\documentclass[onecolumn]{aastex631}
\usepackage{mathtools}
\usepackage{amsmath}
\usepackage{soul}
\usepackage{lipsum}
\usepackage{gensymb}

\newcommand\kms{$\mathrm{km\ s^{-1}}$}
\newcommand\Rsun{$\mathrm{R_{\odot}}$}
\newcommand\Rss{$\mathrm{R_{ss}}$}
\newcommand\car{$\mathrm{C^{6+}/C^{4+}}$}
\newcommand\carr{$\mathrm{C^{6+}/C^{5+}}$}
\newcommand\oxy{$\mathrm{O^{7+}/O^{6+}}$}
\newcommand\feo{$\mathrm{Fe/O}$}

\newcommand\vap{$\mathrm{v_{\alpha p}}$}
\newcommand\sigmac{$\mathrm{\sigma_C}$}
\newcommand\sigmar{$\mathrm{\sigma_R}$}

\newcommand\alf{Alfv\'en}
\newcommand\alfic{Alfv\'enic}
\newcommand\alfty{Alfv\'enicity}

\newcommand\codebank{\url{https://www.github.com/tamarervin/publications/e11_ch}}

\newcommand\rcol{grey}

\newcommand\fieldsdata{\protect\footnote[1]{\url{https://fields.ssl.berkeley.edu/data/}}}
\newcommand\sweapdata{\protect\footnote[2]{\url{http://sweap.cfa.harvard.edu/Data.html}}}
\newcommand\sodata{\protect\footnote[3]{\url{https://soar.esac.esa.int/soar/}}}
\newcommand\gongdata{\protect\footnote[4]{\url{https://gong.nso.edu/}}}
\newcommand\sdodata{\protect\footnote[5]{\url{http://jsoc.stanford.edu/}}}
\newcommand\gongmaps{\protect\footnote[6]{\url{https://gong.nso.edu/adapt/maps}}}

\newcommand\mascode{\protect\footnote[8]{\url{https://www.predsci.com/mhdweb/home.php}}}

\shorttitle{Composition metrics of CH wind}
\shortauthors{Ervin et al.}

\graphicspath{{./}{figures}}

\begin{document}

\title{Compositional metrics of fast and slow {\alfic} solar wind emerging from coronal holes and their boundaries}

\author[0000-0002-8475-8606]{Tamar Ervin}
\affiliation{Department of Physics, University of California, Berkeley, Berkeley, CA 94720-7300, USA; tamarervin@berkeley.edu}
\affiliation{Space Sciences Laboratory, University of California, Berkeley, CA 94720-7450, USA}

\author[0000-0002-1989-3596]{Stuart D. Bale}
\affiliation{Department of Physics, University of California, Berkeley, Berkeley, CA 94720-7300, USA; tamarervin@berkeley.edu}
\affiliation{Space Sciences Laboratory, University of California, Berkeley, CA 94720-7450, USA}

\author[0000-0002-6145-436X]{Samuel T. Badman}
\affiliation{Center for Astrophysics $\vert$ Harvard \& Smithsonian, 60 Garden Street, Cambridge, MA 02138, USA}

\author[0000-0002-8748-2123]{Yeimy J. Rivera}
\affiliation{Center for Astrophysics $\vert$ Harvard \& Smithsonian, 60 Garden Street, Cambridge, MA 02138, USA}

\author[0000-0002-4559-2199]{Orlando Romeo}
\affiliation{Department of Earth and Planetary Science, University of California, Berkeley, CA 94720, USA}
\affiliation{Space Sciences Laboratory, University of California, Berkeley, CA 94720-7450, USA}

\author[0000-0002-9954-4707]{Jia Huang}
\affiliation{Space Sciences Laboratory, University of California, Berkeley, CA 94720-7450, USA}

\author[0000-0002-1859-456X]{Pete Riley}
\affiliation{Predictive Science Inc., San Diego, CA 92121, USA}

\author[0000-0002-4625-3332]{Trevor A. Bowen}
\affiliation{Space Sciences Laboratory, University of California, Berkeley, CA 94720-7450, USA}

\author[0000-0003-1611-227X]{Susan T. Lepri}
\affiliation{Department of Climate and Space Sciences and Engineering, University of Michigan, Ann Arbor, MI 48109, USA}

\author[0000-0003-4437-0698]{Ryan M. Dewey}
\affiliation{Department of Climate and Space Sciences and Engineering, University of Michigan, Ann Arbor, MI 48109, USA}

\begin{abstract}
We seek to understand the composition and variability of fast (FSW) and slow {\alfic} solar wind (SASW) emerging from coronal holes (CH). We leverage an opportune conjunction between Solar Orbiter and Parker Solar Probe (PSP) during PSP Encounter 11 to include compositional diagnostics from the Solar Orbiter heavy ion sensor (HIS) as these variations provide crucial insights into the origin and nature of the solar wind. We use Potential Field Source Surface (PFSS) and Magnetohydrodynamic (MHD) models to connect the observed plasma at PSP and Solar Orbiter to its origin footpoint in the photosphere, and compare these results with the in situ measurements. A very clear signature of a heliospheric current sheet (HCS) crossing as evidenced by enhancements in low FIP elements, ion charge state ratios, proton density, low-{\alfty}, and polarity estimates validates the combination of modeling, data, and mapping. We identify two FSW streams emerging from small equatorial coronal holes (CH) with low ion charge state ratios, low FIP bias, high-{\alfty}, and low footpoint brightness, yet anomalously low alpha particle abundance for both streams. We identify high-{\alfty} slow solar wind emerging from the over-expanded boundary of a CH having intermediate alpha abundance, high-{\alfty}, and dips in ion charge state ratios corresponding to CH boundaries. Through this comprehensive analysis, we highlight the power of multi-instrument conjunction studies in assessing the sources of the solar wind.
\end{abstract}

\section{Introduction} \label{sec:intro}

The solar wind is a stream of ionized plasma (composed of protons, electrons, and alphas along with trace amounts of heavy ions) that continuously escapes from the Sun and interacts with everything in its path \citep{Parker-1958}. While the parameters of the solar wind, such as density, velocity, and ion and elemental composition are highly variable, we can categorize the solar wind into distinct types. The main categorization is set using speed where slow and fast wind have a canonical cutoff between the two of 500 {\kms} at 1AU \citep{Zurbuchen-2007, Borovsky-2012, Stakhiv-2015}, however, there are other characteristics which can be used to differentiate types of solar wind: e.g. {\alfty} \citep{DAmicis-2015, Stansby-2020alf}, particle densities \citep{Mostafavi-2022}, and heavy ion composition \citep{Widing-2001, Zhao-2009, Stansby-2020comp}. While we have decades of measurements of solar wind parameters, we still lack a complete understanding of the acceleration, heating mechanisms, and source regions of the solar wind \citep{Abbo-2016, Cranmer-2019, Viall-2020}. A better understanding of these mechanisms will come from making an effective connection between measurements taken in the heliosphere and relating them to their source region. Ideally, we aim to grasp the complete narrative spanning from the conditions of origination in the corona, through the subsequent evolution, to the moment when spacecraft intercept them in the heliosphere to create a more holistic picture of solar activity and the underlying mechanisms driving the solar wind.

It has been well established through studying Ulysses data covering a large range of heliographic latitudes that during solar minimum high-speed wind primarily originates from polar coronal holes \citep{McComas-1998, McComas-2008}. \citet{vonSteiger-2000} confirmed that faster wind had a more photosphere-like composition in comparison to slow wind, which supported the idea that elemental composition could be used to determine whether wind originated from a coronal hole (CH), active region, or streamer structures at the Sun. While fast solar wind is known to originate from CHs, the origins of the slow solar wind are not as well understood \citep{Viall-2020}. At solar minimum, slow wind has been observed primarily near the heliospheric current sheet \citep{Bavassano-1997, Smith-1978}, while during solar maximum it has been shown to be more interspersed with fast streams \citep{McComas-2001, McComas-2008}. The slow solar wind typically shows more variability (e.g. density, temperature, and chemical makeup) in comparison to the more homogeneous fast solar wind (FSW) \citep{Bruno-1986, Lopez-1986, Schwenn-2006}, exhibiting larger proton density, higher electron temperature, enhanced charge state ratios (\carr, \oxy) and enhancement of elements of low first ionization potential (FIP) \citep[and references therein]{vonSteiger-2000, Abbo-2016}. 

The large variability observed in the parameters of the slow wind suggests that it originates from a diverse set of source regions, likely influenced by position within the solar cycle, as different regions emerge at various stages. \citet{DAmicis-2015, Stansby-2018, Stansby-2020alf} and others have shown the existence of two types of slow solar wind: the classical slow solar wind (SSW) with slow speeds and low-{\alfty}, and the highly {\alfic} slow solar wind (SASW) which shows properties more similar to that of the FSW. Studies using in situ measurements have found that some slow solar wind shows plasma signatures that are very similar to that of the FSW -- high alpha particle abundances \citep{Ohmi-2004}, large differential velocities \citep{Stansby-2020comp}, lower heavy ion charge state ratios and reduced low-FIP enhanced elemental abundances \citep{Stakhiv-2015, DAmicis-2018}, and high {\alfty} \citep{DAmicis-2015, Perrone-2020, DAmicis-2021}. These are typical characteristics of the FSW, which means that it is likely that at least some of the slow solar wind originates from CHs, particularly small CHs or the boundaries of larger ones such that the coronal magnetic field strongly over expands \cite{Wang-1990}.

A major difference in terms of the type of CH region expected to produce fast versus slow wind is the point at which the wind becomes supersonic (the critical point). Heating above this point causes increases in wind speed \citep{Leer-1980}, and since rapidly diverging fields have higher critical points, they produce slower speed winds \citep{Wang-2012, Stansby-2020alf}. Since small CHs have higher magnetic expansion, they are more likely to produce SSW due to the point of energy deposit in the magnetic structure \citep{Nolte-1976, Garton-2018}. This overall correspondence between magnetic expansion and solar wind speed as been well established empirically \citep{Wang-1990} and forms the basis of the Wang-Sheeley-Arge solar wind prediction model \citep{Arge-2000, Arge-2003, Arge-2004}. 

The solar wind is primarily comprised of ionized hydrogen (H) and helium (He), i.e. protons and alpha particles. Their properties, such as velocity, density, and temperature, vary and alpha particles are found at much lower number densities when compared with protons \citep{Bame-1997}. The enhancement or depletion in the alpha-to-proton abundance ratio is characteristic of FSW and HCS crossings, respectively. \citet{Suess-2009} confirms the existence of alpha particle depletions up to 10 days in width and show that, alongside \citet{Borrini-1981, Gosling-1981}, this depletion originates from helmet streamers in closed coronal loops due to transient plasma release from streamer cores which are then sheared due to velocity differences between the plasma from then two streamer legs \citep{Suess-2009}. At solar minimum, the alpha abundance ratio shows significant positive correlation with wind speed, while at solar maximum the correlation is much weaker and converges to the fast wind ratio (4-5\%), which is unchanged through the solar cycle \citep{Kasper-2007, Kasper-2012}. This is interpreted as showing that low alpha abundance slow solar wind comes from the HCS and helmet streamers, which are much more frequently sampled at solar minimum. \citet{Kasper-2016, Alterman-2018, Alterman-2019} have shown that the combination of \lq{}alpha particle rich\rq{} and \lq{}poor\rq{} populations in the slow solar wind thus likely originate from solar minimum helmet streamers (alpha particle poor) and active regions (alpha particle rich).

In addition to hydrogen and helium, there are also small amounts of heavier ions (Z$>$2) in the solar wind \citep{Bame-1997, Bochsler-2007}. In the photosphere, hydrogen has an abundance of 12.00 while helium is at 10.914 $\pm$ 0.013, oxygen (O) at 8.69 $\pm$ 0.04, and iron (Fe) at 7.46 $\pm$ 0.04 \citep{Asplund-2021} where the abundance is $\log \epsilon_X = \log(N_X/N_H) + 12$. We use the photospheric elemental composition values for individual elements as context to understand how abundance values change elsewhere in the solar atmosphere \citep{Asplund-2009}. The processes that ionize and fractionate the heavy ions lead to composition variations, which can be measured both via remote and in situ methods and occur either in the corona or chromosphere. Therefore, the composition of the solar wind is dependent upon the processes at work in the transition layers in the upper solar atmosphere \citep{Laming-2019}. In this work, we look at the {\feo} ratio as a measure of fractionation (changes in elemental abundances summed over charge states) and FIP bias, and the ion charge state ratios of {\oxy} and {\car} as a measure of ionization level. The ion charge ratios are useful for studying the properties of the plasma from which the solar wind originates due to the \lq{}freeze in\rq{} process whereby the ionization state remains constant beyond a certain distance from the solar surface, known as the freeze in point. This occurs because ionization and recombination rates are proportional to the electron density, which rapidly decreases with radial distance from the solar surface until the density is so low that neither of these processes can occur \citep{Owocki-1983}. These ratios are then \lq{}frozen in\rq{}, at a freeze in point of 1.0-1.9{\Rsun} \citep{Buergi-1986, Chen-2003, Landi-2012b}, such that they can be used to determine information about the properties of the coronal plasma from which the wind originates. 

{\car} and {\oxy} ratios are mainly set by the electron source temperature along with some (quite small) radial evolution effects upon reaching an open magnetic field line \citep{Hundhausen-1968}. Higher ion charge ratios are correlated with higher temperature source regions in the corona \citep{Owens-2018}. Thus, low {\oxy} ratios imply solar wind originating from cold CH regions \citep[and references therein]{vonSteiger-2011}, however high {\oxy} ratios have been shown to originate from multiple sources: active regions \citep{Liewer-2003, Ko-2006, Ko-2014, Culhane-2014}, CH boundaries \citep{Antiochos-2012, Crooker-2012}, and the helmet streamer stalk \citep{Zhao-2009, Zhao-2014, Zhao-2011}. \citet{Wang-2016} show that high {\oxy} ratios are correlated with flux tubes emerging from the inner boundaries of CHs and that at 1AU, a threshold value of 0.145 separates CH wind of low {\oxy} ratio from streamer wind (high {\oxy} ratio). While this threshold was determined using measurements at 1AU, it can be applied to measurements closer to the Sun as the freeze-in of ionization states leads to minimal radial propagation effects on this ratio. Both the {\car} and {\oxy} ratios are expected to follow the same trends as they have similar freeze-in heights where the theoretically predicted temperature is monotonically increasing until its maximum, and thus, we expect the charge state ratios to be correlated. However, \citet{Zhao-2017} discuss a type of \lq{}Outlier\rq{} slow solar wind with {\carr} ratios much lower than expected that must originate at 1.6{\Rsun} or higher through an episodic generating process.

In addition to ion charge state ratios, studying FIP bias through the {\feo} ratio provides an additional tracer of coronal source region for comparison with model results. The FIP effect, in which the abundance of lower FIP elements (e.g. Fe) are observed to be enhanced relative to the abundance of higher FIP elements (e.g. O),  was first measured in the solar wind and Solar Energetic Particles (SEPs), then later from spectroscopic coronal observations \citep{Meyer-1985, Bochsler-1986, Gloeckler-1989, Feldman-1992}. Remote measurements of coronal loops show that they exhibit FIP bias that increases the longer the loops exist \citep{Feldman-1997}, and since some slow wind exhibits this type of bias, this indicates that this slow wind likely originates from large loops. In CH regions, we see continuously open field lines, while in other regions, material is trapped and released from large loops that sometimes connect with open field lines. Therefore, coronal material escaping from structures such as streamers or active regions is distinct from material originating from CHs. 



In the low chromosphere, the leading theory is that the majority of MHD waves that accelerate the solar wind are generated from convection-driven motions as pressure changes from thermal to magnetically dominated \citep{Laming-2019}. In the upper chromosphere, neutral gas becomes ionized plasma leading to strong density gradients, which cause {\alf} wave refraction and reflection. This generates the ponderomotive force, which stems from the combination of reflection and refraction of {\alf} waves on plasma ions. Due to the electromagnetic nature of these waves, only ions feel this force. The time average of this force leads to an enhancement of ions over neutrals as it is directed upwards in the solar chromosphere \citep{Laming-2017}, which gives way to ion-neutral separation and the elemental fractionation (the FIP effect) in the upper atmosphere. 

CHs undergo quasi-rigid rotation as the coronal field is continuously undergoing reconnection to remain close to a potential state \citep{Lionello-2005, Wang-2009}. This continuous magnetic reconnection is thought to contribute to to outwardly propagating MHD waves and heating that occurs far above the transition region. In the large loops that contribute to the slow solar wind, MHD waves remain trapped in the transition region, within regions of rapidly diverging open flux where heating is concentrated at low heights \citep{Wang-2009a}, thus heating the ions leading to enhanced charge state ratios. Additionally, the timescale for fractionation is typically 2 to 3 days, meaning plasma must remain in the corona for this time period to experience fractionation \citep{Laming-2015}. This is why plasma from open field lines that quickly escapes the corona does not see an enhancement in low-FIP abundances, while plasma on closed loops (e.g. those found in active regions forming the streamer belt) does \citep{Geiss-1995, vonSteiger-2000}. We use the {\feo} measurement in conjunction with {\alfty}, modeling efforts, and other in situ measurements to provide more powerful diagnostics of different types of solar wind and their source region.

Leveraging a unique conjunction opportunity between inner heliospheric space missions, we use a combination of modeling methods and in situ diagnostics to characterize both FSW and SASW emerging from CHs. We begin with a discussion of the spacecraft and remote sensing data that is used in this study in Section~\ref{sec: data}. We outline the advancements of Parker Solar Probe (PSP; \citet{Fox-2016}) and Solar Orbiter (\citet{Muller-2020}) that allow for this study and how modeling methods are used in conjunction with in situ data to study the source regions of the solar wind. In Section~\ref{sec: methods}, we discuss the two major magnetic field modeling methods used, potential field source surface (PFSS) and magnetohydrodynamic (MHD) models, and how these methods allow us to study the large-scale coronal magnetic field and identify solar wind source regions. We discuss the observations in Section~\ref{sec: obs/disc} specifically looking at elemental composition (Section~\ref{sec: elements}), particle velocity and abundance ratios (Section~\ref{sec: particles}), and modeling efforts (Section~\ref{sec: models}). Lastly, in Section~\ref{sec: results}, we discuss the main results of the study, then finish with our conclusions and ideas for further work in Section~\ref{sec: conclusion}. 


\section{Data} \label{sec: data}

Our study uses a combination of in situ data from PSP \citep{Fox-2016} and Solar Orbiter \citep{Muller-2020} along with magnetograms from the Helioseismic and Magnetic Imager (HMI; \citet{Scherrer-2012}) aboard the Solar Dynamics Observatory (SDO; \citet{Pesnell-2012}) and the Global Oscillation Network Group (GONG; \citet{Harvey-1996}) during the 11th solar encounter of PSP (E11) in February 18 to March 4, 2022. Through novel measurements, orbital trajectories and instrumentation methods, these instruments have revolutionized our study and understanding of the Sun. The use of both PSP and Solar Orbiter data is required as Solar Orbiter provides elemental abundance measurements (such as the {\feo} ratio) and charge state ratios ({\car} and {\oxy}), which PSP does not, and PSP provides a much better quality calculation of {\alfty} (see Appendix~\ref{sec: appendix-sigma}). Additionally, as PSP measures solar wind velocities all below the canonical 500 {\kms} cutoff during this time period, Solar Orbiter velocity measurements (which are closer to the asymptotic 1AU values) are necessary to characterize the wind as fast or slow according to the canonical 1AU cutoff. Lastly, measuring and identifying features at two substantially different heliocentric distances adds robustness to the analysis by verifying the relevant streams are long-lived and not strongly affected by transients or time evolution.


The PSP mission consists of four instruments to study coronal heating, particle acceleration, and energy flow through the corona. From the Electromagnetic Fields Investigation (FIELDS; \citet{Bale-2016}), we use DC magnetic field measurements to understand the global heliospheric field and to validate our models and spacecraft alignments. Proton and alpha particle measurements (density and velocity) from PSP come from the Ion Solar Probe ANalyzer (SPAN-I; \citet{Livi-2022}) on the \lq{}Solar Wind Electrons, Alphas, and Protons\rq{} (SWEAP; \citet{Kasper-2016}) investigation. The core electron density was estimated by fitting a bi-Maxwellian distribution along all pitch angles using SPAN-E 3D electron velocity distribution functions (VDFs) \citep{Whittlesey-2020}, following a similar method by \citet{Romeo-2023}. Typically, electron density measurements are calculated from the quasithermal noise spectrum of measurements taken by the Radio Frequency Spectrometer on PSP \citep{Pulupa-2016}, however, the instrument was not functioning for a large portion of E11, and thus this measurement is unusable for this time period. The proton and alpha data are cleaned based on the field of view (FOV) of the instrument and through comparison with electron density measurements. All E11 data used in this study is publicly available for download from the PSP/FIELDS {\fieldsdata} and PSP/SWEAP data archives {\sweapdata}.

The Solar Orbiter mission has a payload of ten in situ and remote sensing instruments. We use Level 2 (science calibrated) DC magnetic field measurements from the Magnetometer (MAG; \citet{Horbury-2020}); and particle and heavy ion measurements from the Solar Wind Analyzer (SWA; \citet{Owen-2020}) suite. The SWA suite takes high cadence measurements of heavy ion abundances and 3D VDFs of electrons, protons, and alpha particles. The combination of sensors allows for the characterization of the ion and electron bulk properties of the solar wind between 0.28 and 1 AU \citep{Owen-2020}. The Heavy Ion Sensor (SWA/HIS; \citet{Livi-2023}) determines abundances of heavy ions, which allows us to determine elemental composition and charge state ratios of the plasma, key measurements not available with PSP. Heavy ion measurements such as {\feo}, {\oxy}, and {\car} ratios are a Level 3 data product from SWA/HIS with a resolution time of 10 minutes \citep{Owen-2020, Livi-2023}. Using SWA/PAS, we take measured proton fluxes and velocities at a 4-second cadence to characterize the bulk plasma properties of the solar wind. All Solar Orbiter data used in this study can be found on the Solar Orbiter Archive {\sodata}.


We use magnetograms from SDO/HMI and GONG as inputs to our modeling methods (described in Section~\ref{sec: methods}). GONG provides full-disk magnetograms every minute with a resolution of 5 arc seconds, which are publicly available through the NSO website {\gongdata}. HMI provides full-disk Doppler velocity measurements, line of sight photospheric magnetic flux, and continuum images at one arc second resolution with a 45-second cadence along with vector magnetic field measurements at 90 or 135-second cadence \citep{Scherrer-2012}. These measurements are produced with the 6177.3nm Fe I line with a resolution of 0.5 arc seconds per pixel, and the solar disk fills almost the entire image frame in HMI data. In addition to HMI, SDO has the Atmospheric Imaging Assembly (AIA; \citet{Lemen-2012}), which provides high-resolution full-disk images of the solar corona and transition region at a 12-second resolution. AIA consists of four telescopes taking images at a variety of wavelengths. To produce high-resolution, full-Sun Carrington EUV maps for comparison with models, we use images from the Fe XII (193 {\AA}) bandpass filter aboard SDO/AIA. All SDO/HMI and SDO/AIA data can be accessed through JSOC {\sdodata}.

\begin{figure} [htb!]
  \includegraphics[width=\columnwidth]{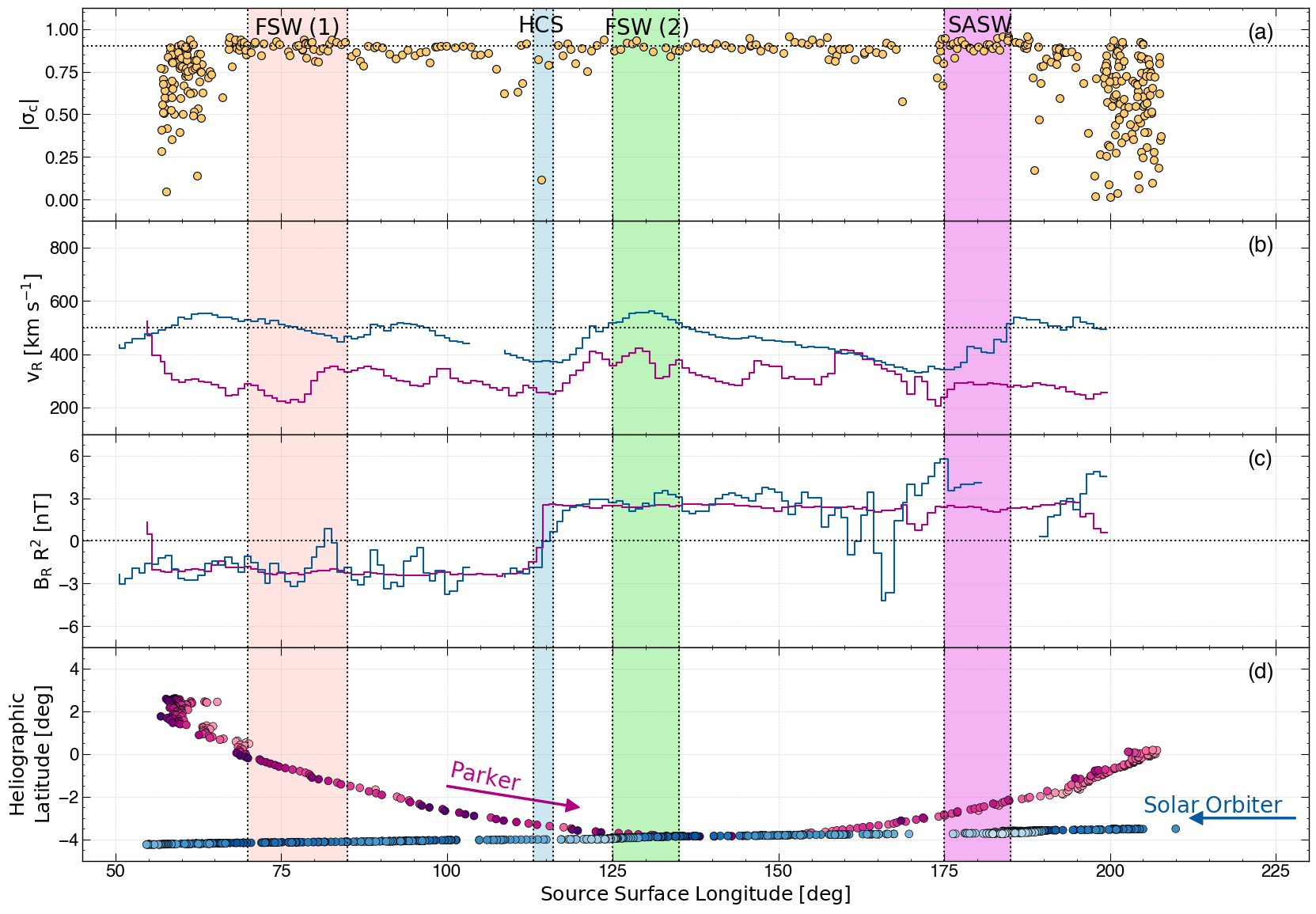}
  \caption{Parker spiral alignment between PSP and Solar Orbiter during PSP E11 showing identification of regions of interest to this study. The HCS in blue, FSW (1) in pink, FSW (2) in green, and SASW in purple. \textit{Panel (a):} Absolute cross helicity (an {\alfty} proxy) as calculated from PSP measurements and outlined in Equation~\ref{eq: sigmac}. The dashed line at 0.9 shows the cutoff between {\alfic} and non-{\alfic} wind. \textit{Panel (b):} Comparison of radial velocity as measured by PSP (pink) and Solar Orbiter (blue). The dashed line shows the canonical 500 {\kms} cutoff between fast and slow wind at 1AU. \textit{Panel (c):} The scaled radial magnetic field as measured by PSP (pink) and Solar Orbiter (blue) shows the alignment of the HCS crossings and the conservation of flux. The dashed line at 0 shows where the spacecraft crossed the neutral line. Velocity and magnetic field measurements are in 1{\degree} longitude bins. \textit{Panel (d):} The trajectories of PSP and Solar Orbiter, in heliographic longitude and latitude, showing the direction of travel and the Parker spiral alignment during this time period. Points are colored by the average solar wind speed in each 1{\degree} longitude bin.}
  \label{fig: alignment}
\end{figure}

In order to carry out a conjunction study using PSP and Solar Orbiter data, we are interested in time periods when both spacecraft are observing the same plasma. These time periods are those when the spacecraft is aligned along a Parker spiral streamline, or measuring the same magnetic footpoint longitude in Carrington coordinates, and ideally at similar latitudes. We identify periods of alignment by mapping the latitude and longitude of the spacecraft's orbit down to a few solar radii. Panel (c) in Figure~\ref{fig: alignment} shows the position of the spacecraft during the perihelion of E11 and their Parker spiral alignment. This time period is ideal for a case study as measurements for all quantities needed in this study were available, and we cross a variety of solar magnetic structures during the spacecraft's trajectory. \citet{Rivera-2023} show evidence of this conjunction for part of this interval through four methods: matching magnetic field polarities, mapping velocity peaks to the same longitudes, and matching helium abundance and mass flux between the spacecraft.

In order to directly compare measurements between PSP and Solar Orbiter, we plot the observations as a function of projected longitude using ballistic propagation (Figure~\ref{fig: alignment}). \citet{Snyder-1966} first proposed this method of connecting an ideal Parker spiral field line from a point in the heliosphere to the corona, assuming a constant solar wind speed. \citet{Nolte-1973} demonstrated the capabilities of this method to provide a reliable estimate of solar surface footpoints. While this technique does not accurately represent the real solar wind, which has a non-constant speed; \citet{Nolte-1973, Macneil-2022, Koukras-2022} estimates the error to be less than 10{\degree} in longitude as the two corrections, coronal corotation and solar wind acceleration, cancel to first order. We propagate field lines inward from their initial position in the PSP and Solar Orbiter trajectory to their associated latitude and longitude on the source surface at 2.0{\Rsun} using a varying solar wind speed (the in situ $\mathrm{v_R}$ measurement) to most accurately represent the real solar wind following \citet{Badman-2020} Equation 1. This allows for direct comparison of observations as a function of source longitude.

In Figure~\ref{fig: alignment}, we show the mapped in situ data and classification of our regions of interest shaded as follows: blue (HCS crossing), pink (FSW period 1), green (FSW period 2), and purple (SASW). Wind is classified as fast or slow using the Solar Orbiter velocity measurement with the 500 {\kms} cutoff (panel (b)), and we identify regions of high (or low) {\alfty} by looking at the absolute cross helicity ($\mathrm{|\sigma_C|}$) and residual energy ($\mathrm{|\sigma_R|}$) as defined in Equations~\ref{eq: sigmac} and~\ref{eq: sigmar} (panel (a)). $\mathrm{|\sigma_C|}$ is a proxy for the {\alfty} of the plasma where higher cross helicity measurements ($\mathrm{|\sigma_C|}$ values of 1 or -1) are indicative of pure {\alf} waves propagating parallel or antiparallel to the magnetic field. Plasma is considered to have high (or low) {\alfty} when $\mathrm{|\sigma_C|}$ is above (or below) 0.9, while the residual energy tells us the difference between the energy in velocity and magnetic field fluctuations and which is dominant in the plasma.

\begin{equation} \label{eq: sigmac}
    \mathrm{\sigma_C = \frac{<E^+> - <E^->}{<E^+> + <E^->} }
\end{equation}

\begin{equation} \label{eq: sigmar}
    \mathrm{\sigma_R = \frac{2 <z^+ \cdot z^->}{<E^+> + <E^->} }
\end{equation}

In these calculations, $<\cdot \cdot \cdot>$ corresponds to a time average over a 20-minute non-overlapping time window, a timescale typical of {\alfic} fluctuations \citep{Tu-1995}. $\mathrm{E ^ \pm}$ is the energy ($\mathrm{E ^ \pm} = |(z ^\pm) ^ 2|$) of the Els{\"a}asser \citep{Elsasser-1950} variables $\mathrm{\mathbf{z} ^ \pm = \delta \mathbf{v} \pm \delta \mathbf{B}/\sqrt{\mu_0 m_p N_p}}$ \citep{Wicks-2013, Chen-2013} where $\mathrm{\delta \mathbf{v}}$ and $\mathrm{\delta \mathbf{B}}$ are the fluctuations of velocity and magnetic field. We use a standard method to validate our cross helicity results in Figure~\ref{fig: mag-cross-helicity} in the Appendix (Section~\ref{sec: appendix-sigma}) by showing the circular shape (as expected) within a radius of 1 of the $\mathrm{\sigma_C}$ versus $\mathrm{\sigma_R}$ (residual energy) plot \citep{Bavassano-1998} and further discuss the degradation of the cross helicity calculation between PSP and Solar Orbiter, showing the necessity of PSP measurements to use this quantity as a classifier. Panel (c) is the scaled radial magnetic field showing flux conservation and the HCS alignment achieved through ballistic propagation. Panel (d) shows the source surface heliographic longitude and latitude, highlighting that the latitudes were nearly identical (a rare occurrence) and that the spacecraft moved over structures in the opposite direction. 

\begin{figure*} [htb!]
  \includegraphics[width=\textwidth]{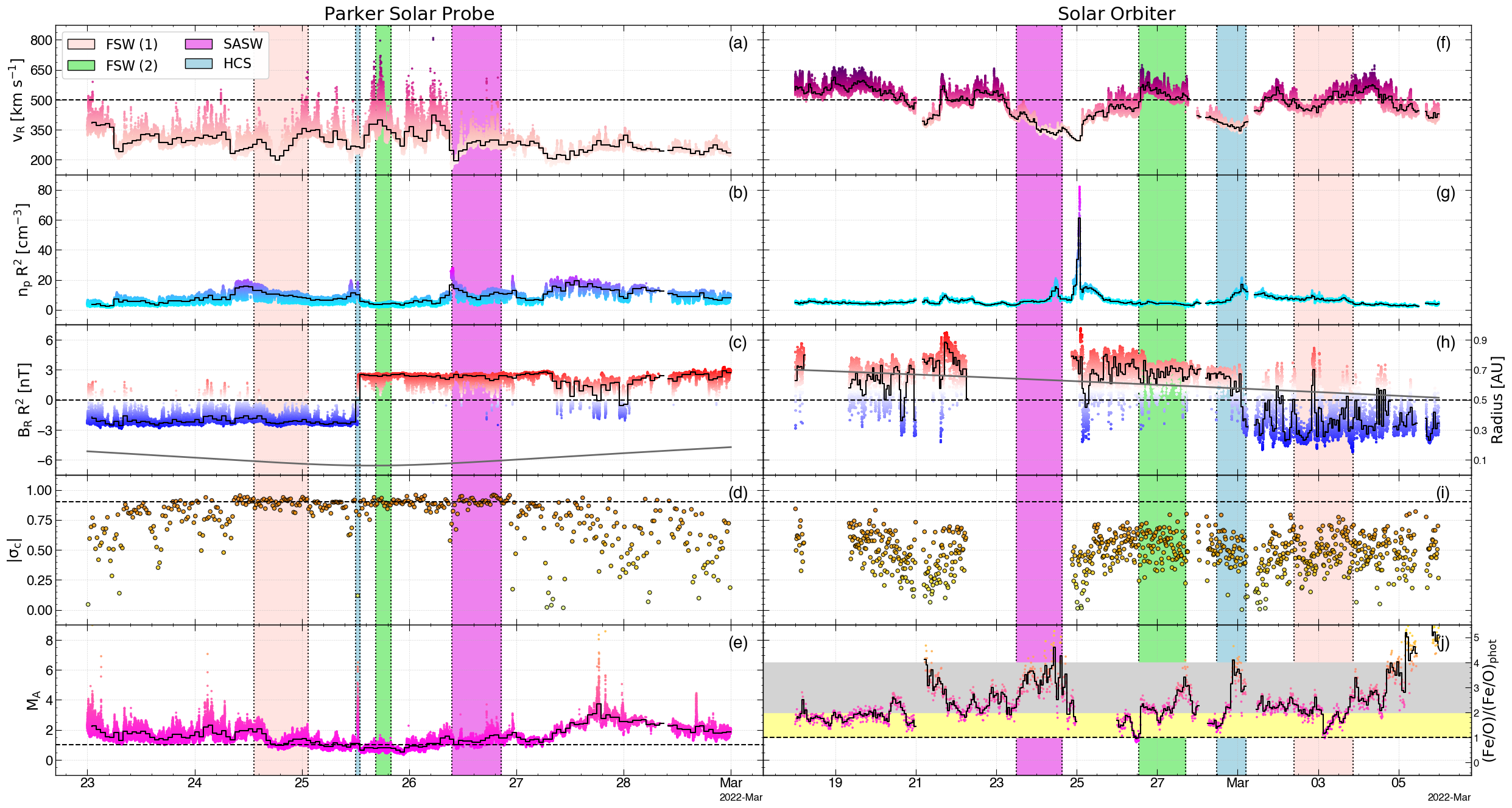}
  \caption{A timeseries showing the data from the period of interest from PSP (left column) and Solar Orbiter (right column). The top panels (panels (a) and (f)) show the solar wind velocity (\kms), the second panels (panels (b) and (g)) show scaled proton density ($\mathrm{cm^{-3}}$), and the third panels (panels (c) and (h)) show the scaled radial magnetic field (nT). We also show the radial position of the spacecraft in AU ({\rcol}) in panels (c) and (h) alongside the scaled radial magnetic field. In Panels (d) and (i), we show the absolute cross helicity ($\mathrm{|\sigma_C|}$) as measured by PSP and Solar Orbiter, along with a dashed line at 0.9 showing the cutoff between {\alfic} and non-{\alfic} wind. The bottom panel in the PSP column (panel (e)) shows the {\alf} Mach number and periods below one are considered sub-{\alfic}. Panel (j) shows the ({\feo})$\mathrm{/(Fe/O)_{phot}}$ as measured by the SWA/HIS instrument aboard Solar Orbiter. This panel is shaded in yellow to show ratios between 1 and 2 (typical of coronal hole wind) and grey to show regions between 2 and 4 (typical of other types of wind). The velocity panels (panels (a) and (f)) include a dashed horizontal line at 500 {\kms} showing the canonical cutoff between fast and slow wind, and the magnetic field data (panels (c) and (h)) has a horizontal dashed line at the neutral point. The data is overlaid with binned data (1-hour cadence) in black. We highlight the regions of interest in time as corresponding to the longitudinal spans highlighted in Figure~\ref{fig: alignment}. These regions correspond to the same position in source surface longitude for both spacecraft when using ballistic propagation to map between the locations of the two spacecraft, but correspond to different periods and lengths of time for PSP and Solar Orbiter due to variances in the direction and speeds of the trajectories.}
  \label{fig: timeseries}
\end{figure*}

We show the full set of the data of interest in Figure~\ref{fig: timeseries} as a timeseries of PSP (left column) and Solar Orbiter (right column) measurements: velocity (panels (a) and (f)), scaled proton density (panels (b) and (g)), and scaled radial magnetic field (panels (c) and (h). Both proton density and magnetic field are scaled by $\mathrm{R^2}$ for the time period of interest. Panels (d) and (i) show the cross helicity ($\mathrm{|\sigma_C|}$) as calculated for both PSP and Solar Orbiter. We see the measurement from Solar Orbiter does not allow us to characterize the {\alfty} of the plasma properly, due to \textbf{gaps in the magnetic field observations (panel (h)) which prevent the calculation of $\mathrm{v_A}$} and the decoherence of the cross helicity measurement with radial distance. Panel (e) shows the Mach number ($\mathrm{M_A}$) from PSP while panel (j) shows the SWA/HIS {\feo} ratio normalized to its photospheric abundance (0.0589) from \citep{Asplund-2021}, a strong tracer of the source region. The {\alfic} Mach number the ratio of the bulk solar wind velocity to the {\alf} speed $\mathrm{(M_A = v_R/v_A)}$. Periods with $\mathrm{M_A}$ below 1 are considered sub-{\alfic}. For this study, we use all the PSP data shown along with the measurements from SWA/HIS on Solar Orbiter and use our ballistic mapping results (Figure~\ref{fig: alignment}) to track the mapped features in the timeseries.

\section{Coronal Modeling} \label{sec: methods}

We extend the backmapped spacecraft locations of the PSP and Solar Orbiter measurements all the way to their estimated photospheric footpoints with PFSS and MHD modeling. We use and compare these modeling methods to each other to provide confidence in our backmapping and footpoint estimations. By comparing these results to several in situ data constraints we evaluate this mapping, learn more about the plasma properties of the source regions and how they evolve from PSP to Solar Orbiter ($\sim$10 to $\sim$120 \Rsun). Both methods are described below. 

\subsection{Potential Field Source Surface Modeling} \label{sec: pfss}

In this work, we use the PFSS model along with an ideal Parker spiral to estimate the photospheric footpoints of the plasma of interest. The PFSS model is an extremely computationally effective method for the determination of the large-scale features of the global coronal magnetic field, typically producing results that compare well with more sophisticated models \citep{Riley-2006, Badman-2023}. This modeling allows for the prediction of results from in situ measurements, and the potential to understand many of the outstanding questions in plasma physics such as coronal heating, and solar wind origins \citep[e.g. ][]{Badman-2020}. 

The PFSS model assumes a magnetostatic (current-free) corona with an inner boundary of direct photospheric magnetograph observations of the radial magnetic field \citep{Schatten-1969, Altschuler-1969} and an outer boundary as an equipotential surface where magnetic field lines are assumed to be open and purely radial such that plasma can escape. This outer boundary has a canonical value of 2.5 \Rsun \citep{Hoeksema-1984}.

We create PFSS models using the \texttt{pfsspy} package, an open-source Python package for Potential Field Source Surface Modeling \citep{pfss}. Using the PFSS modeling method, we trace magnetic field lines from a uniform grid on the photosphere out to the radial source surface. As inputs to the PFSS model, we use synoptic radial magnetic field maps and a value for the source surface height at which the field is set to be fully radial (\Rss). The \texttt{pfsspy} package creates a full 3D magnetic field within this low coronal domain along with the ability to trace individual field lines from a photospheric footpoint.

For the input synoptic radial magnetic field map, we use magnetograms produced using the Air Force Data Assimilative Photospheric Flux Transport (ADAPT; \citet{Worden-2000, Arge-2010, Arge-2011, Arge-2013, Hickmann-2015}) model. This model uses flux transport processes to create a more accurate picture of the global photospheric magnetic field by modeling far-side evolution. ADAPT magnetograms are produced using magnetograms from both HMI and GONG. In this work, we use ADAPT-GONG magnetograms downloaded from the National Solar Observatory (NSO) archive{\gongmaps} and choose a radial source surface height of 2.0 {\Rsun} to complete the PFSS boundary conditions. This is determined to be the optimal value by varying the source surface height until we find the strongest matching between modeled and detected coronal holes and between the modeled and observed HCS \citep{Badman-2022}. This modeling traces magnetic field lines between the source surface at 2.0 {\Rsun} and the solar photosphere, but PSP and Solar Orbiter make measurements around 13.3 {\Rsun} and 120 {\Rsun} respectively during this time period. To fill this gap, we use a ballistic propagation model as discussed in Section~\ref{sec: data} to propagate field lines inward and connect the spacecraft trajectories to the PFSS model. This combination then allows us to estimate the photospheric footpoints from which the plasma emerged. We show a comparison of footpoint estimations for varying source surface heights in panel (d) of Figure~\ref{fig: mhd-comp}, with the black footpoints (2.0 {\Rsun}) as the footpoints we use for this study. In Appendix~\ref{sec: appendix-mag}, we vary the velocity used for ballistic propagation by $\pm$10 {\kms} and the source surface location of PSP by $\pm 5^{\circ}$ to understand how small errors in propagation can effect our estimated footpoints. We find that variance in both source surface height and inducing small errors into the ballistic propagation do not change the resulting source region of the observed wind.

\subsection{Magnetohydrodynamic Models} \label{sec: mhd}

In addition to our PFSS modeling technique, we use an MHD model to compare with and validate our PFSS results. The MHD equations are a set of coupled differential equations for large-scale electrically conducting fluids. We use the Predictive Science Inc. (PSI) Magnetohydrodynamic Algorithm outside a Sphere (MAS) code to solve the MHD equations on a non-uniform mesh grid \citep{Riley-2021}{\mascode}. The MAS code is driven by an observed photospheric radial magnetic field measurement from either SDO/HMI or GONG as an inner boundary condition for the solar magnetic field at 1 {\Rsun} and a heating mechanism. In this paper, we use the semi-empirical thermodynamic solution \citep{Lionello-2001, Lionello-2009}, which solves the MHD equations with empirical thermodynamic approximations of energy transport. MAS models are run on two regimes, the coronal regime from 1 {\Rsun} to 30 {\Rsun} and the heliospheric regime from 30 {\Rsun} out to 1 AU, which improves the efficiency of the computation. The model calculates observables such as velocity, density, and radial magnetic field, which can be compared with in situ and ground-based observations. 

\begin{figure*} [htb!]
  \includegraphics[width=\textwidth]{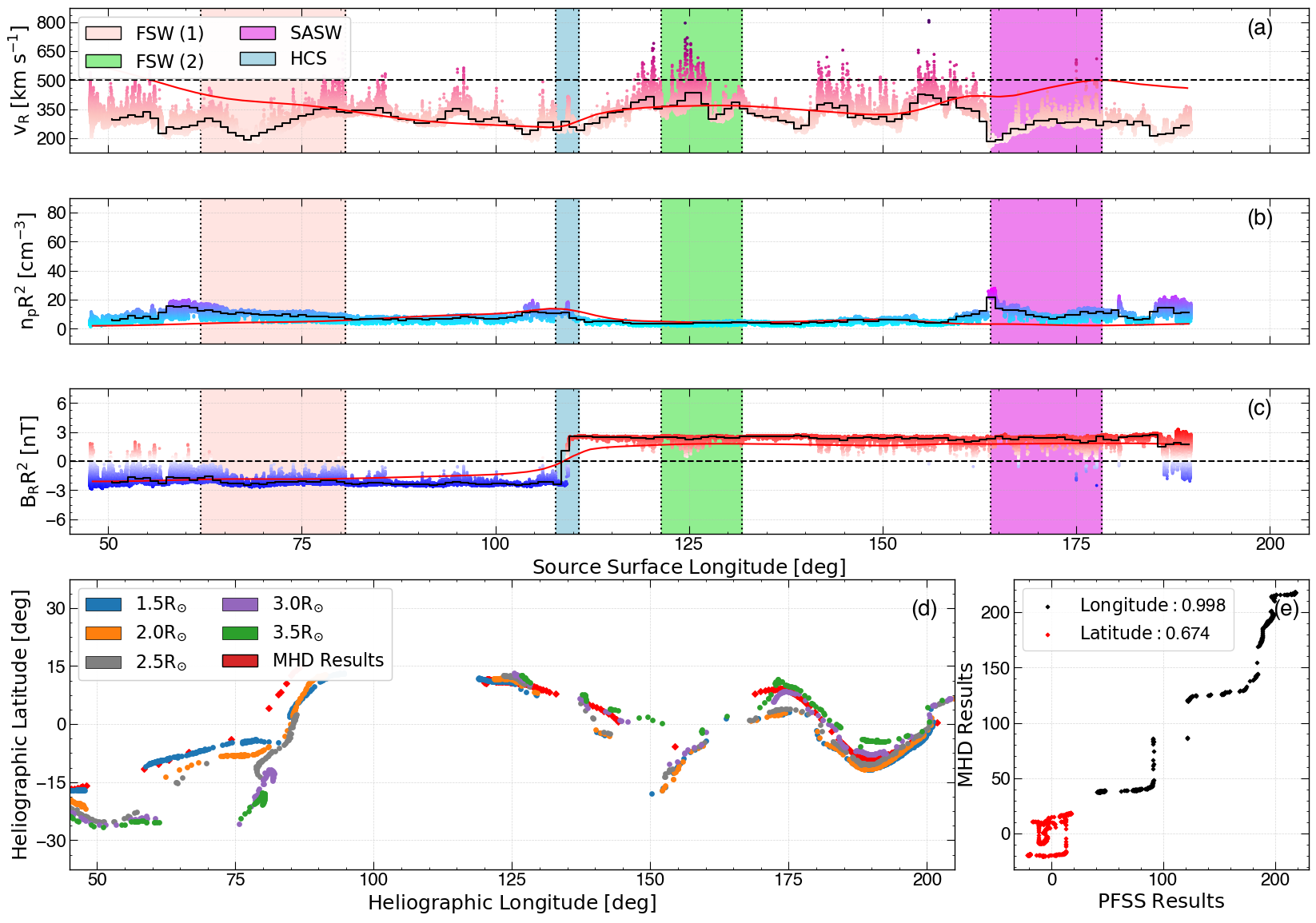}
  \caption{MHD model produced with the PSI MAS code \citep{Riley-2021} for PSP E11 using a SDO/HMI magnetogram from February 24, 2022 as the input boundary condition. The top three panels compare the observables as modeled by the coronal MHD solution (red) at the location of PSP throughout its trajectory, with in-situ measurements of the same parameters from PSP. We show the raw data from PSP overlaid with data binned by 1 degree in longitude (black). \textit{Panel (a):} Comparison between modeled and in-situ radial proton velocity. \textit{Panel (b):} Comparison between the scaled proton density from the MHD model and the PSP measurement. \textit{Panel (c):} Modeled scaled radial magnetic field (red) compared with the measured scaled radial field (black). \textit{Panel (d):} Comparison of the photospheric footpoints between the MHD (red) and PFSS model solutions. We show estimated footpoints from PFSS models at different {\Rss} heights. \textit{Panel (e):} The correlation between the two models with coefficients of 0.99 and 0.62 for longitude and latitude, respectively.}
  \label{fig: mhd-comp}
\end{figure*}

In order to directly compare the measured observables to the model, we sample the MHD modeled observables of interest at PSP's location \textit{in 3D} for the full trajectory (Figure~\ref{fig: mhd-comp}) and compare with the radial magnetic field, density, and velocity measurements from PSP which are taken fully within the coronal model regime. A full Carrington map of the radial cuts of the magnetic field, density, and velocity observables compared with in-situ measurements can be found in the Appendix (see Section~\ref{sec: appendix-mag} Figure~\ref{fig: mhd}). We find very good agreement between the measured and modeled observables, validating that our model is properly functioning and is usable to trace magnetic field lines and estimate photospheric footpoints. In Figure~\ref{fig: mhd-comp}, we show the correlation between the PFSS and MHD estimated footpoints: 0.998 in longitude and 0.674 in latitude for our chosen model with a source surface height of 2.0{\Rsun}. The MHD tracing has no assumptions of ballistic propagation or source surface height and thus provides a strong comparison point with our estimated PFSS footpoints. This, coupled with the strength of the MHD model in reproducing measured plasma parameters, allows for confidence in the footpoints as estimated via the PFSS and MHD models.  

\section{Observations and Discussion} \label{sec: obs/disc}

We combine data from both PSP and Solar Orbiter during PSP E11 with modeling results to \textbf{characterize the plasma emerging from FSW and SASW}. In Figure~\ref{fig: timeseries}, we see that both spacecraft had sharp HCS crossings that map very closely to each other (Figure~\ref{fig: alignment}) while traveling in opposite directions in the solar corotating frame: PSP on February 25, 2022 and Solar Orbiter on March 1, 2022. The HCS crossing is sharper in the PSP measurements as PSP moves at a much larger tangential speed than Solar Orbiter. 

We use ballistic mapping to map the observed plasma back to the source surface and plot the data as a function of Carrington longitude rather than as a timeseries. This allows us to more easily directly compare structures observed by both spacecraft and relate SWA/HIS composition measurements to PSP in situ signatures. In this section, we outline the source regions of the streams of interest and then describe the associated in situ characteristics. 

\subsection{Modeling} \label{sec: models}

\begin{figure*} [htb!]
  \includegraphics[width=\textwidth]{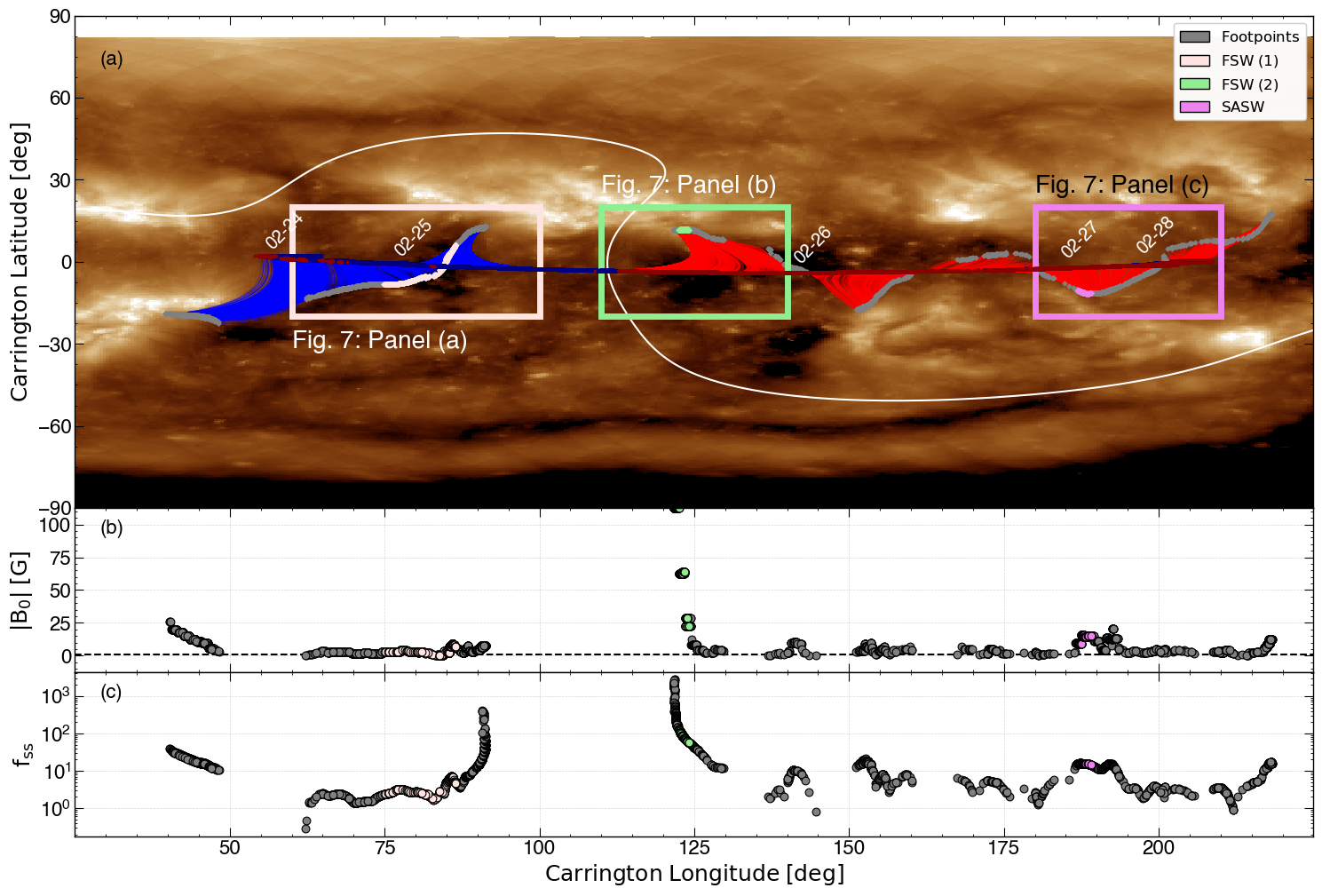}
  \caption{PFSS model for PSP E11 using an input ADAPT-GONG magnetogram from February 24, 2022. \textit{Panel (a):} In this panel, we show a full-Sun Carrington map produced using SDO/AIA 193{\AA} images and overlay the results from our PFSS solution on top. The modeled HCS is in white and the trajectory of the spacecraft is shown colored by the polarity of the magnetic field. The pink, green, and purple rectangles outline the regions of the estimated footpoints of the FSW streams and the SASW stream, which are shown in more detail in Figure~\ref{fig: region}. \textit{Panel (b):} The absolute photospheric magnetic field strength (G) at the estimated footpoints as a function of footpoint longitude. \textit{Panel (c):} The expansion factor calculated following the method of \citep{Wang-1997}.}
  \label{fig: pfss}
\end{figure*}

Using the PFSS and MHD modeling methods, we can estimate the footpoints of the plasma measured in situ by PSP and Solar Orbiter. Our model is able to connect 77\% of the observed plasma to a photospheric source footpoint. In Figure~\ref{fig: mhd-comp}, we show that the predicted observables ($\mathrm{v_{sw}}, \; \mathrm{B_r}, \; \mathrm{N_p}$) from the MHD model explain the PSP in situ measurements well. This coupled with the strong agreement between the PFSS and MHD footpoint estimation gives us confidence in the location of the photospheric footpoints we use for analysis. In Figure~\ref{fig: pfss}, we show the PFSS model for PSP E11 using an ADAPT-GONG magnetogram as the lower boundary input and a source surface height of 2 \Rsun. We show the modeled HCS along with the footpoints produced using a combination of PFSS modeling and ballistic propagation as described in Section~\ref{sec: pfss}. PFSS models extract coronal holes through connecting open field lines to their photospheric footpoints. The model allows us to understand the large-scale structure of the corona, showing a vertical HCS warp, negative polarity near equatorial coronal holes to the left of the HCS, and positive polarity near the equatorial holes to the right of the HCS that is consistent with the measured polarity from PSP and Solar Orbiter. The vertical HCS warp was also seen by \citet{Liewer-2023} in PSP/WISPR data, providing further validation of our model. The modeled HCS in both the MHD and PFSS models shows great agreement with the measured radial magnetic field polarity from both spacecraft, which suggests that the ballistic propagation method works well and the coronal model is accurate, meaning our footpoints are well estimated.

\begin{figure*} [htb!]
  \includegraphics[width=\textwidth]{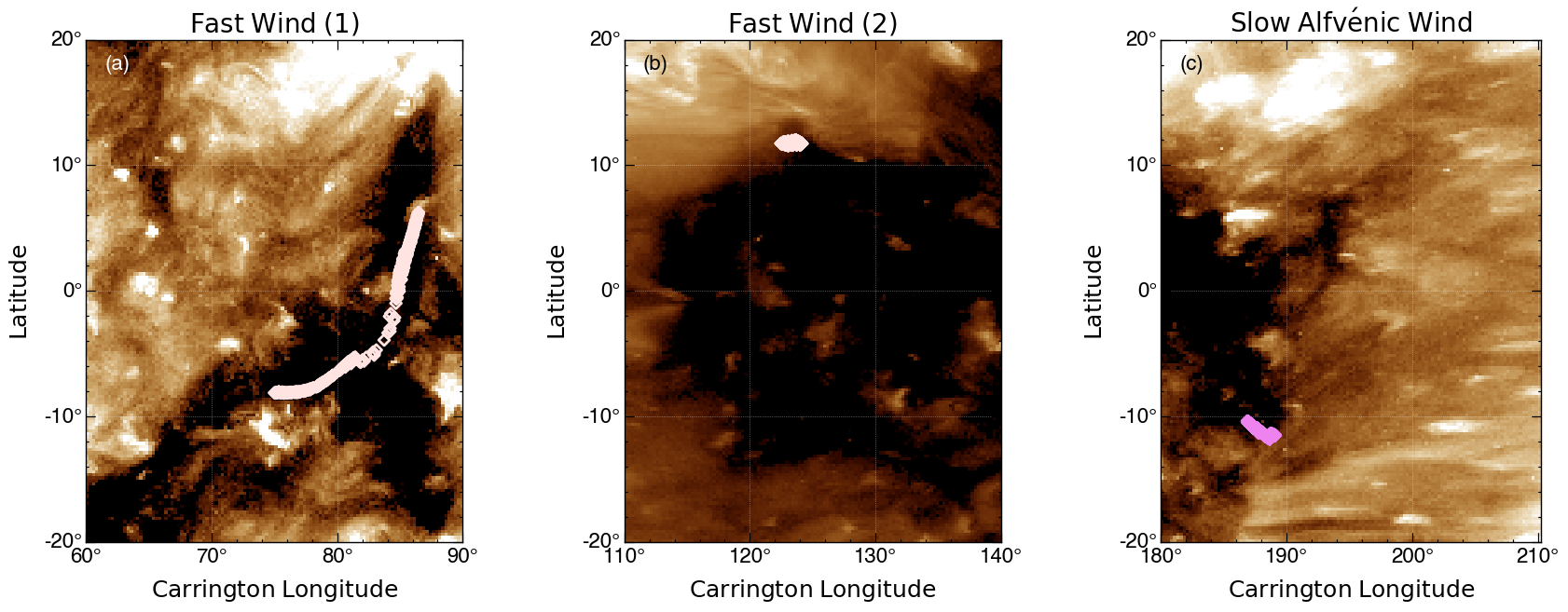}
  \caption{A zoomed in view of the source region connection for the FSW and SASW streams of interest. \textit{Panel (a):} Estimated photospheric footpoints for the FSW (1) stream (pink diamonds) overlaid on a SDO/AIA image of the region. This region corresponds to the pink box in panel (a) of Figure~\ref{fig: pfss}. 
  \textit{Panel (b):} The estimated photospheric footpoints from the FSW (2) stream (green diamonds) showing the connection to the center of a equatorial CH. This region corresponds to the green box in panel (a) of Figure~\ref{fig: pfss}.
  \textit{Panel (c):} The estimated photospheric footpoints from the SASW stream (purple diamonds) showing the connection to the edge of a equatorial CH. This region corresponds to the purple box in panel (a) of Figure~\ref{fig: pfss}.}
  \label{fig: region}
\end{figure*}

Figure~\ref{fig: pfss} allows us to understand the large-scale magnetic field configuration during this time period. To better understand the specific source regions of each wind stream, we look at a combination of modeling results. Figure~\ref{fig: region} shows the source regions for the three interesting periods. In panels (a) and (b), we overlay the estimated footpoints of the FSW and SASW streams on SDO/AIA images. The images are from a time estimate of when the observed wind left the solar surface, which we calculate using the measured wind speed and the location of the spacecraft. This allows for the strongest comparison with the conditions at the time, such that we are better able to understand the source region.

We see that the FSW (pink footpoints) measured by PSP originates from a coronal hole region around 80{\degree} in Carrington longitude. The footpoints at the center of the FSW region are in darker (cooler) coronal plasma, which is consistent with the observation of a dip in the {\oxy} and {\car} ratios near the fast wind region as seen in Figure~\ref{fig: abundance} which corresponds to lower electron temperatures. The footpoints nicely trace the shape of the CH and provide strong justification that this fast wind stream emerged from an equatorial coronal hole. From Figure~\ref{fig: pfss}, we also see low photospheric magnetic field strength and low relative footpoint brightness for this FSW region, both characteristics of wind emerging from CHs. This result also provides additional validation of our footpoint estimations, as it aligns with the expectation for FSW streams.

\textbf{The second FSW stream (green footpoints) emerges from a CH region near 125{\degree} in longitude. Due to the proximity of this FSW stream to the HCS in the model, the expansion factor and footpoint field strength are much larger than for the other FSW (pink) region. In Figure~\ref{fig: region}, we see these footpoints lie within a dark region in SDO/AIA EUV observations, which combined with the low charge state and FIP bias of this period, indicates CH wind.}

The SASW stream shows some similar properties to the FSW stream. In panel (c) of Figure~\ref{fig: region}, we show the SASW stream emerging from the boundary of an equatorial CH. This period has an elevated expansion factor (panel (c) of Figure~\ref{fig: pfss}), which is characteristic of wind streaming from the over-expanded CH boundary \citep{PV-2013, Panasenco-2013, Sheeley-2013, Wang-2013, Panasenco-2019}.

\subsection{Heavy Ion and Elemental Composition Signatures} \label{sec: elements}


By combining the elemental composition and charge state measurements from Solar Orbiter (measurements PSP does not provide) with PSP measurements close to the Sun, which degrade with radial distance (such as {\alfty}), we can more effectively relate heliospheric structure to their coronal sources.

\begin{figure} [htb!]
  \includegraphics[width=\columnwidth]{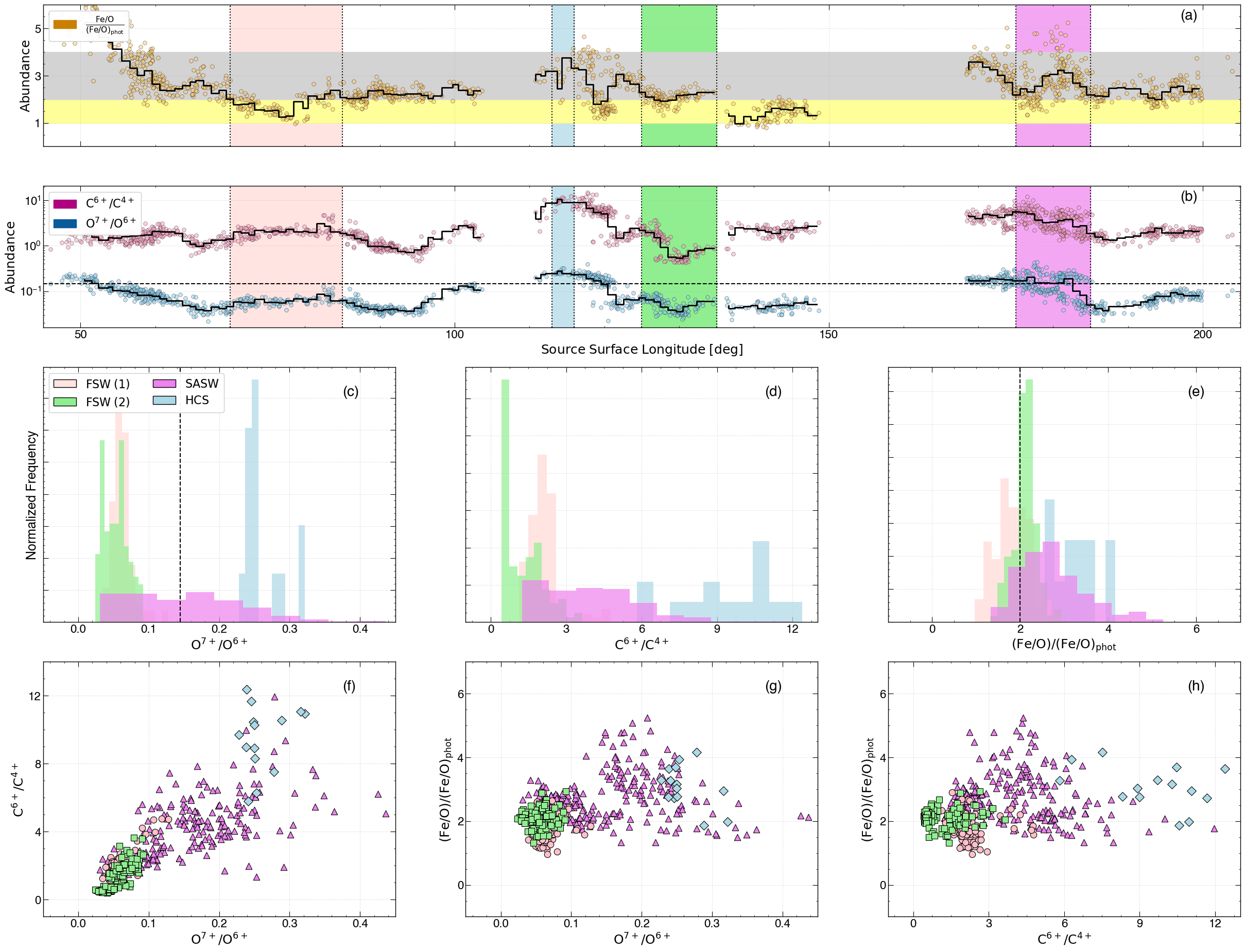}
  \caption{
  \textit{Panel (a):} The normalized {\feo} ratio as per \citet{Asplund-2021} as a function of source surface longitude. We shade regions where the abundance ratios are between 1 and 2 in yellow, and between 2 and 4 in grey. We overplot the raw data with binned data at a 1{\degree} longitude cadence. \textit{Panel (b):}  Ion charge state ratios from SWA/HIS overplotted with 1{\degree} in longitude binned data. We include a dashed vertical line at 0.145, which is shown to be the threshold for the {\oxy} ratio separating coronal hole wind from streamer wind \citep{Wang-2016}. 
  \textbf{\textit{Panels (c), (d), and (e):}  1D normalized histograms of each of the compositional metrics, colored by wind type. In panel (c), we include a dashed vertical line at 0.145, which is shown to be the threshold for the {\oxy} ratio separating coronal hole wind from streamer wind \citep{Wang-2016}. The dashed line in panel (c) indicates the typical upper threshold for photospheric FIP abundance levels. \textit{Panels (f), (g), and (h):} Comparison between {\oxy} and {\car}, {\oxy} and {\feo}, and {\car} and {\feo} ratios respectively for the regions of interest. Correlations for each wind regime can be found in Table~\ref{tab: his-corr}.}}
  \label{fig: abundance}
\end{figure}

\textbf{In Figure~\ref{fig: abundance}, we outline the ion ({\car}, {\oxy}) and elemental ({\feo}) composition measurements.} The {\feo} ratio is scaled to its photospheric value according to ((\feo)/$\mathrm{(Fe/O)_{phot}}$) where $\mathrm{(Fe/O)_{phot}}$ = 0.0589 \citep{Asplund-2021}. From this point on, discussion of the {\feo} ratio will refer to the normalized ((\feo)/$\mathrm{(Fe/O)_{phot}}$) value. The normalized {\feo} ratio has been shown to relate to coronal source region properties and alongside modeling is a good tracer of solar wind source. Ratios between 1 and 2 are indicative of photosphere-like composition typical of CH regions, while ratios from 2 to 4 show coronal-like composition typical of non-CH wind. Similarly, the ion charge state ratios in panels (a) and (b) are a diagnostic of a coronal temperature, allowing us to probe the thermal properties of the stream's coronal source. We include a dashed line in panel (a) at 0.145 which is the cutoff for the {\oxy} ratio from \citet{Wang-2016} between streamer (high {\oxy}) and CH wind.


We see variance in the {\feo} ratio for all wind types, which implies a combination of source regions for the plasma that is measured throughout the encounter. We primarily see ratios ranging between 2 and 4, which is usually the case for active region outflow as found in \citet{Feldman-2000, Widing-2001}. The majority of this time period consists of low {\car} and {\oxy} ratios, which imply a low characteristic electron source temperature. Throughout the encounter, the {\oxy} ratio varies between 0.022 and 0.44, while the {\car} ratio shows a far wider variance between 0.42 and 14.9. The {\feo} ratio ranges between 0.80 and 5.25 for this time period.


In panels (d), (e), and (f) of Figure~\ref{fig: abundance} and in Table~\ref{tab: his-corr}, we investigate the correlations between the charge state and elemental abundance ratios. The data points are colorized according to the time interval of interest shown in the left-hand panel. The ion ratios ({\car} and {\oxy}) are strongly correlated with each other, with a Spearman correlation coefficient of 0.75, but not with the {\feo} ratio (Figure~\ref{fig: abundance}). Both the {\car} and {\oxy} ratios are indicators of electron temperature with some radial evolution effects and as such, are expected to follow the same trends as they have similar freeze-in heights \citep{Zhao-2017}. The {\car} and {\oxy} ratios have Spearman correlation coefficients of 0.15 and 0.40 with {\feo}, respectively. The lack of overall correlation between the charge state ratios and elemental composition implies the presence of a variety of solar wind sources. There has not been a strong correlation observed between {\feo} and solar wind speed as seen by both SWA/HIS and ACE/SWICS \citep{Livi-2023} and so we would not necessarily expect a strong correlation with the charge state ratios and {\feo}. This could be because the characteristic source region temperature can be transient in nature (heating and cooling occurring at noticeable timescales) while the FIP effect takes place on a longer timescale, or the outflow effects are important -- the density, temperature, and bulk speed that set the ion ratios during outflow varies across source regions with similar {\feo}. 

When looking at the correlations for the different time periods of interest, we see some distinct groupings. The FSW streams typically shows far lower ion ratios than the other types of wind which is expected for CH wind. It shows a strong correlation between {\oxy} and {\car} but is relatively uncorrelated in the other panels. Similar to the FSW, the SASW has a higher correlation between the charge state ratios. Looking at the correlation plots, we see two groups of SASW, most distinct in panel (d). In all three panels, there is a lower left blob of points more similar to FSW-like characteristics and an upper right blob of points found nearer to the HCS wind. 



\begin{table}[ht]
\centering
    \begin{tabular}{|c|c|c|c|}
    \hline
    \textbf{Wind Type} & {\oxy} vs. {\car} & {\oxy} vs. {\feo} & {\car} vs. {\feo} \\
    \hline
    \textbf{All} & 0.75 & 0.40 & 0.15 \\
    \hline
    \textbf{HCS} & 0.05 & -0.24 & -0.40 \\
    \hline
    \textbf{FSW (1)} & 0.54 & 0.02 & -0.35 \\
    \hline
    \textbf{FSW (2)} & 0.84 & 0.038 & -0.006 \\
    \hline
    \textbf{SASW} & 0.71 & 0.16 & -0.08 \\
    \hline
    \end{tabular}
    \caption{Spearman correlation coefficients between charge state ratios ({\oxy} and {\car}) and {\feo}, a measure of the low FIP bias in the solar wind. We show the correlation for each type of wind along with the correlation for the entire time period.}
\label{tab: his-corr}
\end{table}

The largest change in the {\car} and {\oxy} ratios occurs at the current sheet crossing, where plasma tends to be slow, dense, and hot, which allows more time for ionization thus producing higher abundances of $\mathrm{O^{7+}}$ and $\mathrm{C^{6+}}$. These sharp increases in ionization state of these species indicate regions of increased coronal electron temperatures. This is a common effect when crossing the HCS and provides a validation metric for comparing our modeling methods with observations. In this encounter, we see an increase of 608\% in the {\car} abundances, 130\% in the {\oxy} abundances, and 113\% in the {\feo} ratios over 3{\degree} in longitude as we cross the HCS. 

We see low {\oxy} and {\car} ratios during the FSW streams, and a dip in these ratios just before and after the regions showing a transition to cooler, less dense coronal hole plasma. The {\feo} ratio (see Figure~\ref{fig: timeseries}) during this both FSW streams has photosphere-like abundance levels with the sharpest dip in the center of the region -- indicative of CH type plasma. \textbf{For the charge state ratios and {\feo} ratio, the FSW streams show incredibly low values indicating that the spacecraft were sampling wind from deep within a CH.}

The SASW region has a huge variance in all ratios. The {\feo} (see Figure~\ref{fig: timeseries}) ratio increases and decreases through the center of the region indicating some variation in the structure the spacecraft is connected to. Both ion charge state ratios show large variation in this region from more CH-type wind to streamer-like wind. \textbf{While the values for the SASW are not as distinctively low as for the FSW streams, they are far below the HCS wind indicating their origin from a cooler source region.} As discussed previously, there seem to be two groupings of composition results for the SASW. This is due to the overlapping in the backmapping in this region, as seen in Figure~\ref{fig: alignment}.



\subsection{Particle Measurements} \label{sec: particles}

The two primary ion components that make up the solar wind are protons and alpha particles, and in this section, we discuss the variance in proton-to-alpha abundance ratio and differential velocity. Similar to the ion ratios discussed in Section~\ref{sec: elements}, the alpha particle abundance ratio ($\mathrm{A_{He} = N_{\alpha}/N_p}$) provides insight to identifying the source regions of the solar wind. The relative velocity of alpha particles to protons can vary depending on the type of solar wind and decreases with radial distance from the Sun, and often shows a positive correlation with the bulk solar wind speed \citep{Mostafavi-2022}. Despite being more massive, alpha particles have higher velocities than protons close to the Sun \citep{Feldman-1997}, an example of preferential acceleration at work.

\begin{figure} [htb!]
  \includegraphics[width=\columnwidth]{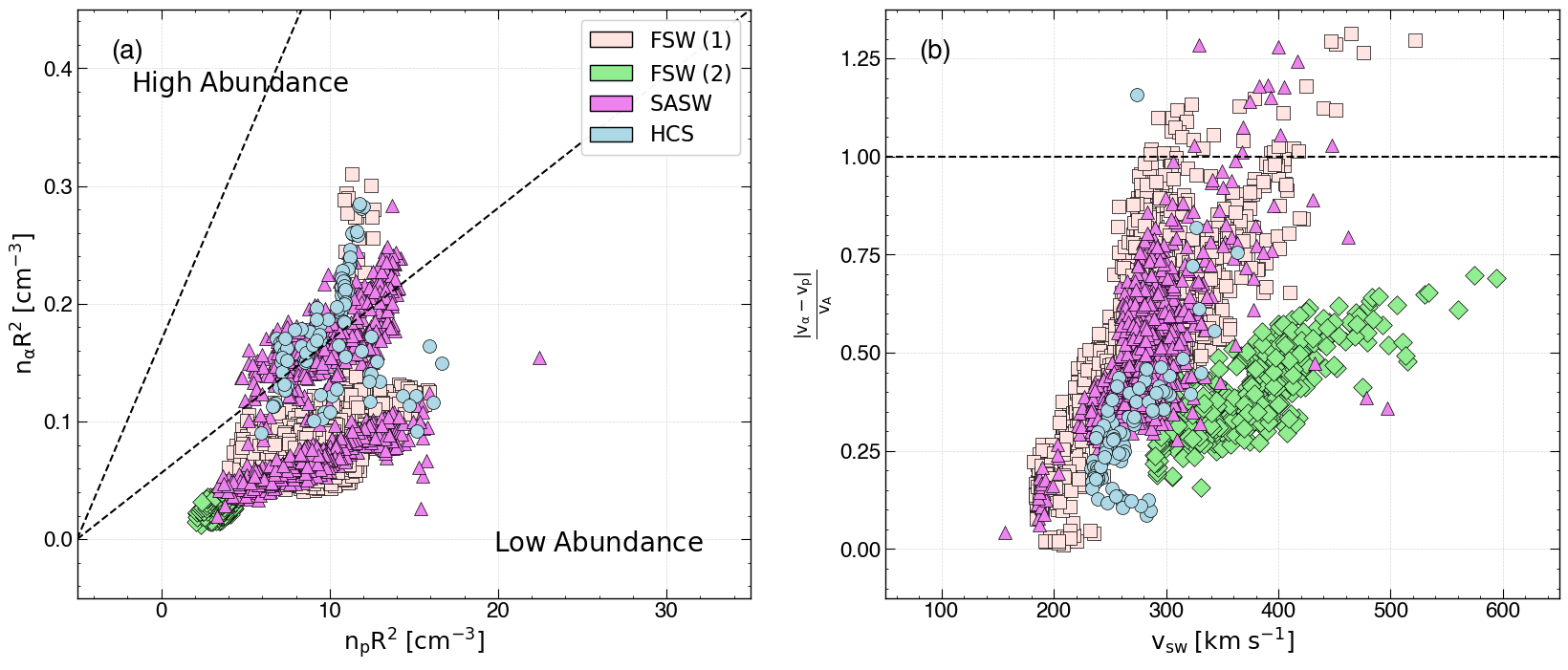}
  \caption{\textbf{Particle measurements from the SPAN-I instrument aboard PSP/SWEAP showing the alpha-to-proton abundance ratio and normalized velocity differential. \emph{Panel (a):} The scaled proton density ($\mathrm{n_p R^2}$) against the scaled alpha particle density ($\mathrm{n_{\alpha} R^2}$) as a function of the four time periods of interest based on their corresponding colors. The dashed lines show the high abundance boundary of 0.045 and the low abundance boundary of 0.015. \emph{Panel (b):} The normalized velocity differential ({\vap}) against the solar wind speed ($\mathrm{v_{sw}}$) as a function of the four time periods of interest based on their corresponding colors. The dashed line at 1 shows when the velocity differential is less than the {\alf} speed.}}
  \label{fig: vel/dens}
\end{figure}

In Figure~\ref{fig: vel/dens} we show the PSP particle measurements. In panel (a), we show the scaled proton and alpha abundance as a function of wind type and in panel (b) the differential velocity normalized by the {\alf} speed ($\mathrm{v_{\alpha p} = \frac{|v_{\alpha} - v_p|}{v_A}}$) as a function of wind speed. The dashed lines in panel (a) separate between the high ($\mathrm{A_{He}} \geq 0.045$) and low ($\mathrm{A_{He}} \leq 0.015$) abundance regimes as defined by \citet{Kasper-2007, Kasper-2012}. 



In Figure~\ref{fig: vel/dens}, we see that overall, the abundance ratio during E11 rarely crosses above the typical \lq{}high abundance\rq{} threshold of 0.045 \citep{Kasper-2007, Kasper-2012}. There is some variance in abundance throughout the encounter, however the ratio remains low for the majority of this period. There is a strong correlation between the proton ($\mathrm{v_p}$) and alpha particle ($\mathrm{v_{\alpha}}$) bulk velocity with a Spearman correlation of 0.9. The correlation between the differential velocity and the solar wind velocity is weaker (correlation of 0.75) and is stronger between the fast solar wind and differential velocity (0.88 and 0.63 for FSW (1) and FSW (2)) than for the SASW (0.59). We find that the alpha-to-proton differential velocity is typically below the {\alf} speed, $\mathrm{v_{\alpha p}/v_A}$ (panel (b)). This is due to alpha-proton instabilities driving wave-particle interactions to hit the instability threshold, leading to an upper bound on the differential speed of $\mathrm{v_{\alpha p}/v_A} \leq 1$ \citep{Gary-2000}. 

At the HCS crossing, we see an increase in proton density in both the PSP and Solar Orbiter data (Figure~\ref{fig: timeseries}) along with a decreased in helium abundance in panel (a) of Figure~\ref{fig: vel/dens}. There are two clusters of abundance ratios: one just above the low abundance threshold and one just below. There is also a wide range of scatter in the differential velocity measurement, with the binned value around 0.25.

In the FSW regions, we see intermediate but primarily very low helium abundances. This is atypical for FSW but is seen in both streams discussed during this study and thus warrants further investigation. The differential velocity correlates well with wind speed as is expected \citep{Mostafavi-2022}.

The SASW consists of both alpha particle intermediate and poor plasma populations implying that it likely originates from multiple source regions. We can see these two groupings in panel (a) where one group shows more HCS-like abundance ratios, and the other shows more FSW-type ratios. It shows differential streaming speeds very similar to the FSW populations and a strong correlation between {\vap} and wind speed.




\section{Results} \label{sec: results}

\textbf{By connecting models to in situ data, we are able to probe the characteristics of the solar wind emerging from CHs.} The connection between modeling and in situ data provides insight from both a theoretical and observational point of view as to the origins of the solar wind. While the modeling methods do have user-defined inputs, when combined with in situ observations, they provide strong evidence to characterize the compositional signature of varying types of CH wind.

\begin{enumerate}

\item Similar to \citet{Chen-2021}, we see non-Alfvenic slow wind emerging from the HCS supported by an enhancement in the proton density, low alpha-to-proton abundance, and high {\feo}, {\oxy}, and {\car} ratios.

  The in situ measurements of the magnetic field from near the HCS crossing provide observational constraints and validation for our PFSS and MHD models while the composition metrics is typical of plasma in closed coronal loops. HCS plasma typically is slow, dense, and hot where higher $\mathrm{O^{7+}}$ and $\mathrm{C^{6+}}$ abundances are often observed. The high {\feo} ratio shows strong low-FIP enhancement has taken place in association with closed loop structures in connection to the streamer belt \citet{Wang-2000b, Howard-2008, Sheeley-2009, Rouillard-2010, Liewer-2023} with helium abundance enhancements near the HCS. We also see a depletion in the helium abundance ratio near the HCS crossing characteristic of streamers in closed coronal loops. These are all typical in situ metrics for classical non-{\alfic} slow wind.

\item As expected from \citet{vonSteiger-2000, McComas-1998, McComas-2008}, we find that two fast wind (FSW) streams that maps to relatively large coronal hole regions, with low ion charge state ratios and photospheric-like Fe/O abundances, characteristic of coronal hole wind.

    As shown in Figure~\ref{fig: abundance}, a relatively lower ({\oxy} and {\car}) ratio in the fast wind stream indicates solar wind originating from a cool CH structure at the Sun. The normalized {\feo} ratios observed in this region fall in the photosphere-like regime, typical of wind emerging from open field lines that has not had the time to fractionate. The modeling results connect the plasma from this region to a large coronal hole with low relative footpoint brightness, providing additional evidence of the source region of the FSW. 

\item The slow {\alfic} wind (SASW) during this time period is shown to have two populations of plasma with different characteristics: one that shows more FSW-type characteristics and the other showing more streamer-like properties. Similar to results from \citet{DAmicis-2021aa}, the SASW is thought to originate from the over-expanded edges of a CH boundary.

The SASW shows coronal-like {\feo} abundance ratios with a spike in the center of the region, showing some variance and perhaps evolution in the structure the spacecraft are magnetically connected to. The ion abundance ratios bound this region with a large range of measured values. The correlation plots show two groupings: one with high charge state ratios and low FIP enhancement and the other with low charge state ratio and photospheric level elemental composition. This regime also shows variance in alpha-to-proton abundance with a grouping similar to the HCS abundance and one similar to the FSW abundance. These groupings along with the modeling results in Figure~\ref{fig: pfss} imply the SASW comes from \textbf{the over-expanded boundary of a} positive polarity coronal hole. Over expansion can occur due to the presence of coronal pseudostreamers, which when coupled with low coronal non-monotonic expansion can slow the solar wind leading to similar properties to the FSW, but much slower wind speeds \citep{PV-2013, Panasenco-2019, DAmicis-2021aa}. 

  
\end{enumerate}

\section{Conclusion} \label{sec: conclusion}

While the source regions of the slow solar wind still remain an open question, the novel in situ measurements that the new generation of spacecraft, PSP and Solar Orbiter, provide allow us to delve deeper into this question and study the nature of the solar wind. Combining the spacecraft instrumentation via conjunctions allows us to make detailed measurements of the solar wind evolution through comparing quantities known to decohere with radial distance, such as the {\alfty}, with radially invariant quantities, such as the elemental composition and charge state ratios, which are fixed in the plasma at low coronal altitudes and therefore utilize the wider range of instrumentation available further from the Sun. In this paper, we outlined the modeling methods and measurements we use to study the coronal source origin of fast and slow {\alfic} solar wind and connect these methods to in situ plasma measurements. With PFSS and MHD modeling, we detect fast wind from deep in coronal holes while the slow {\alfic} solar wind originates from an \textbf{over-expanded} CH boundary. 

We then look at elemental abundances and particle measurements to study the FIP effect and trace coronal source region. In situ measurements from PSP and Solar Orbiter support our modeling results: fast solar wind has low FIP bias as measured by {\feo} (Figure~\ref{fig: abundance}), very low ion charge state ratios implying a relatively cool coronal source, and higher alpha particle abundance ratios in comparison with the slow wind region, characteristic of wind streaming along open field lines from coronal holes. The slow {\alfic} wind shows two plasma populations, one with FSW properties (low FIP, low charge state ratios) and another more similar to classical slow wind (higher charge state, higher FIP bias). In addition to providing evidence towards the source region of the plasma, the in situ measurements provide interesting insight into the plasma escaping from higher in the corona. There is a clear enhancement in charge state ratios and FIP bias as the spacecraft cross the HCS. This characteristic HCS signature seen in the SWA/HIS data shows that we are able to attach composition information to a PSP data set, extending its science potential and capabilities.






While the combination of modeling methods and novel in situ measurements provide an incomparable insight into the nature and source regions of the solar wind, there is still much work to do to accurately and convincingly determine the source regions of the solar wind (especially the slow solar wind). Modeling methods are limited by the assumptions and input data that we choose to use for our specific study and more work is necessary to create quantitative methods to determine the best input data to use for a model. For example with the PFSS model, the choice of input magnetogram and source surface height drastically impacts the location and shape of the modeled HCS and in situ measurements are required to determine which model most accurately describes the large-scale structure of the coronal magnetic field. These types of models are also hindered by time and spatial evolution of magnetic structures on the solar surface. While the ADAPT flux transport models account for some of the temporal and spatial evolution, there is still more work to be done to create accurate global models of the evolution of the solar surface and more spacecraft to fly to allow for multiple viewpoints of images at once. This conjunction occurred while the source regions were behind the limb, and models would be more robust if we had full-Sun magnetograph coverage. 

We see very low alpha abundance in the FSW region and additional exploration of PSP fast wind periods should be done to understand if this is an effect of PSP's radial position, this encounter specifically, or some other mechanism at work. The slow wind showed a large spread in source properties and coronal conditions for similar solar wind speed, from a priori very similar looking corona, and more work looking at the characteristics of the slow wind in comparison to its {\alfty} and source region must be done to completely characterize the slow wind and its origination mechanisms. Additionally, further work exploring the impact of source region on wind temperature and the radial evolution of the parameters discussed in this paper (velocity, density, charge state ratios, FIP bias, helium abundance, {\alfty}, etc.) should be done to fully characterize the solar wind propagating from various source regions. These types of studies require the existence and identification of future multi-spacecraft conjunctions to uncover the effects of radial propagation and have access to a variety of remote and in situ measurements to fully understand the processes at work creating and driving the solar wind.

\section{Acknowledgements} \label{sec: acknowledgements}

The authors would like to thank the reviewer for their suggestions and feedback throughout the review process.

The FIELDS and SWEAP experiments on the PSP spacecraft was designed and developed under NASA contract NNN06AA01C. PR gratefully acknowledges support from NASA (80NSSC20K0695, 80NSSC20K1285, 80NSSC23K0258), and the Parker Solar Probe WISPR contract NNG11EK11I to NRL (under subcontract N00173-19-C-2003 to PSI).

We acknowledge the NASA Parker Solar Probe Mission and the SWEAP team led by J. Kasper for use of data.

Solar Orbiter is a mission of international cooperation between ESA and NASA, operated by ESA. Funding for SwRI was provided by NASA contract NNG10EK25C. Funding for the University of Michigan was provided through SwRI subcontract A99201MO. R.M.D. acknowledges support from NASA grant 80NSSC22K0204.


This work utilizes data produced collaboratively between Air Force Research Laboratory (AFRL) \& the National Solar Observatory (NSO). The ADAPT model development is supported by AFRL. The input data utilized by ADAPT is obtained by NSO/NISP (NSO Integrated Synoptic Program). NSO is operated by the Association of Universities for Research in Astronomy (AURA), Inc., under a cooperative agreement with the National Science Foundation (NSF).

This work utilizes GONG data obtained by the NSO Integrated Synoptic Program, managed by the National Solar Observatory, which is operated by the Association of Universities for Research in Astronomy (AURA), Inc. under a cooperative agreement with the National Science Foundation and with contribution from the National Oceanic and Atmospheric Administration. The GONG network of instruments is hosted by the Big Bear Solar Observatory, High Altitude Observatory, Learmonth Solar Observatory, Udaipur Solar Observatory, Instituto de Astrofísica de Canarias, and Cerro Tololo Interamerican Observatory.

SDO/AIA and SDO/HMI data is courtesy of NASA/SDO and the AIA, EVE, and HMI science teams.

This research used version 4.1.6 of the SunPy open source software package \citep{sunpy}, and made use of HelioPy, a community-developed Python package for space physics \citep{heliopy}. All code to replicate figures can be found at {\codebank}.

\software{
\texttt{Astropy} \citep{astropy:2013, astropy:2018, astropy:2022},
\texttt{heliopy} \citep{heliopy},
\texttt{matplotlib} \citep{mpl},
\texttt{numpy} \citep{numpy},
\texttt{pandas} \citep{pandas},
\texttt{psipy} \citep{psipy},
\texttt{pfsspy} \citep{pfss},
\texttt{scipy} \citep{scipy},
\texttt{spiceypy}\citep{spiceypy},
\texttt{SunPy} \citep{sunpy}
}

\section{Appendix A: {\alfty}} \label{sec: appendix-sigma}
\begin{figure*} [htb!]
  \includegraphics[width=\textwidth]{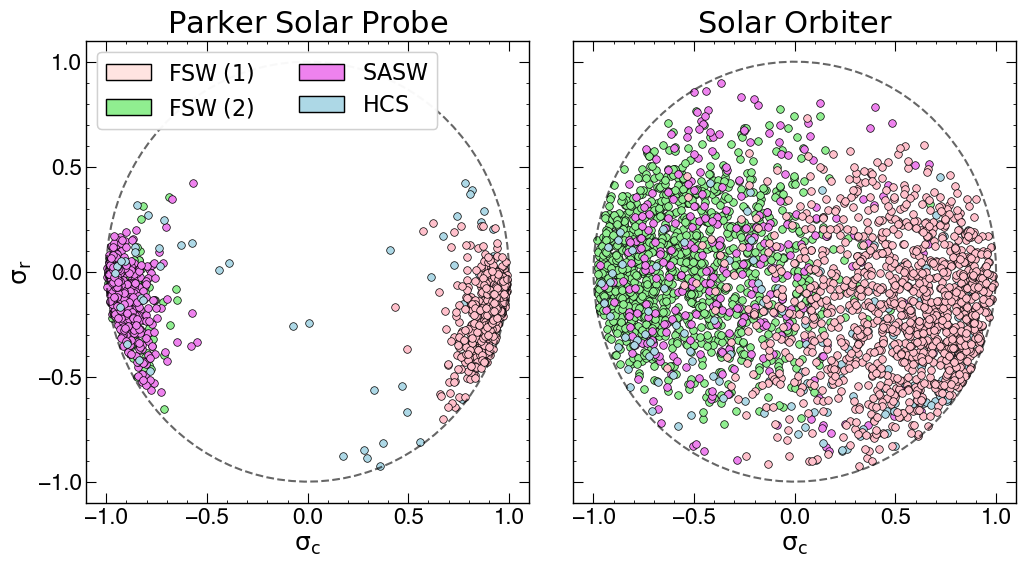}
  \caption{Comparison of the cross helicity ($\mathrm{\sigma_C}$) and residual energy ($\mathrm{\sigma_R}$). Points are colored based on the region of interest they correspond to. SSW is in green, SASW is in purple, FSW is in pink, and HCS wind is in blue. The dotted circle signifies the region within which we expect points to lie as $\mathrm{\sigma_C^2 + \sigma_R^2 \leq 1}$ \citep{Bavassano-1998}.}
  \label{fig: mag-cross-helicity}
\end{figure*}

We validate our cross helicity and residual energy results by looking at Figure~\ref{fig: mag-cross-helicity}. The cross helicity gives information about the level of {\alfic} fluctuations in the plasma while the residual energy is the difference between the magnetic and kinetic energy. We expect a circular shape when the cross helicity ($\mathrm{\sigma_C}$) is plotted against the residual energy ($\mathrm{\sigma_R}$), which we see in Figure~\ref{fig: mag-cross-helicity} and is a validation of the cross helicity calculation we used to determine {\alfty} of the plasma.

Figure~\ref{fig: mag-cross-helicity} also shows the degradation of the measure of cross helicity as a tool to quantify {\alfty} between PSP and Solar Orbiter. In the PSP measurements, we see the HCS plasma with a {\sigmac} near 0 and {\sigmar} values near -1. We expect HCS plasma to have low {\alfty} ({\sigmac} $\sim$ 0) as it is a source of SSW and {\sigmar} $\sim$ -1 implies it is strongly dominated by magnetic field fluctuations. The SASW (purple) shows high levels of {\alfty} in the PSP data ({\sigmac} $\sim$ -1) while in the Solar Orbiter data, it has a cross helicity value closer to 0. The FSW shows less differentiation between the PSP and Solar Orbiter measurements, in both cases with high {\sigmac} and {\sigmar} values near 0. The high {\alfty} populations show {\sigmar} values near 0, meaning there is a rough balance between magnetic and kinetic energy, implying nearly pure {\alf} fluctuations close to the Sun \citep{Kasper-2019}. Overall, we see that the cross helicity measure becomes erased further from the Sun, proving the necessity of using both PSP and Solar Orbiter measurements to study this time period. This also shows some of the radial propagation effects of cross helicity and should be studied more fully with future multi-spacecraft conjunctions.

\section{Appendix B: Magnetic Field Modeling} \label{sec: appendix-mag}

\begin{figure*} [htb!]
\centering
  \includegraphics[width=0.55\textwidth]{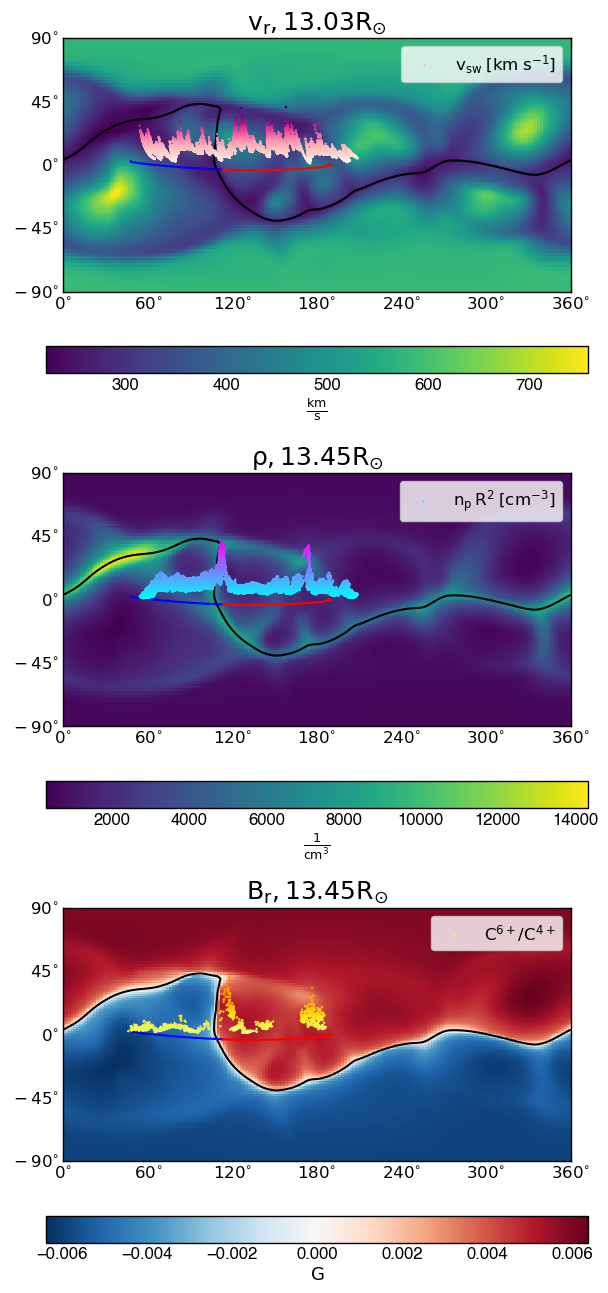}
  \caption{Radial cuts of the MAS MHD model \citep{Riley-2021} for PSP E11 using a SDO/HMI magnetogram from February 24, 2022 as the input boundary condition. The top panel shows the modeled radial velocity, middle panel shows the modeled density and the bottom panel shows the modeled radial magnetic field. We overlay the trajectory of PSP during E11 color-coded by the measured magnetic field polarity. The neutral line from the model is shown in black. We include in situ measurements of solar wind velocity, scaled proton density, and the radial magnetic field from PSP/FIELDS.}
  \label{fig: mhd}
\end{figure*}

Our work relies upon two primary methods for modeling the solar coronal magnetic field. Both PFSS and MHD models have been shown to reproduce observed magnetic fields \citep{Badman-2020} and are a trusted method for determining the source origin of observed plasma. In Figure~\ref{fig: mhd}, we show a comparison of the radial cuts of the MHD model at 13.3{\Rsun} (closest radial cut of the model to the PSP perihelion distance), as discussed in section~\ref{sec: mhd}, with in-situ measurements. 

The top panel shows a radial cut at 13.03{\Rsun} of the radial velocity as modeled by MAS, along with the modeled HCS (in black). We see regions of faster wind emerging from between streamer arcs. In the middle panel, we show the radial cut of the density at 13.45{\Rsun} with density enhancements in the model corresponding to the HCS and streamer regions. In the bottom panel, we compare the model radial magnetic field at 13.45{\Rsun} with the measurements from PSP/FIELDS. 

\begin{figure*} [htb!]
  \includegraphics[width=\textwidth]{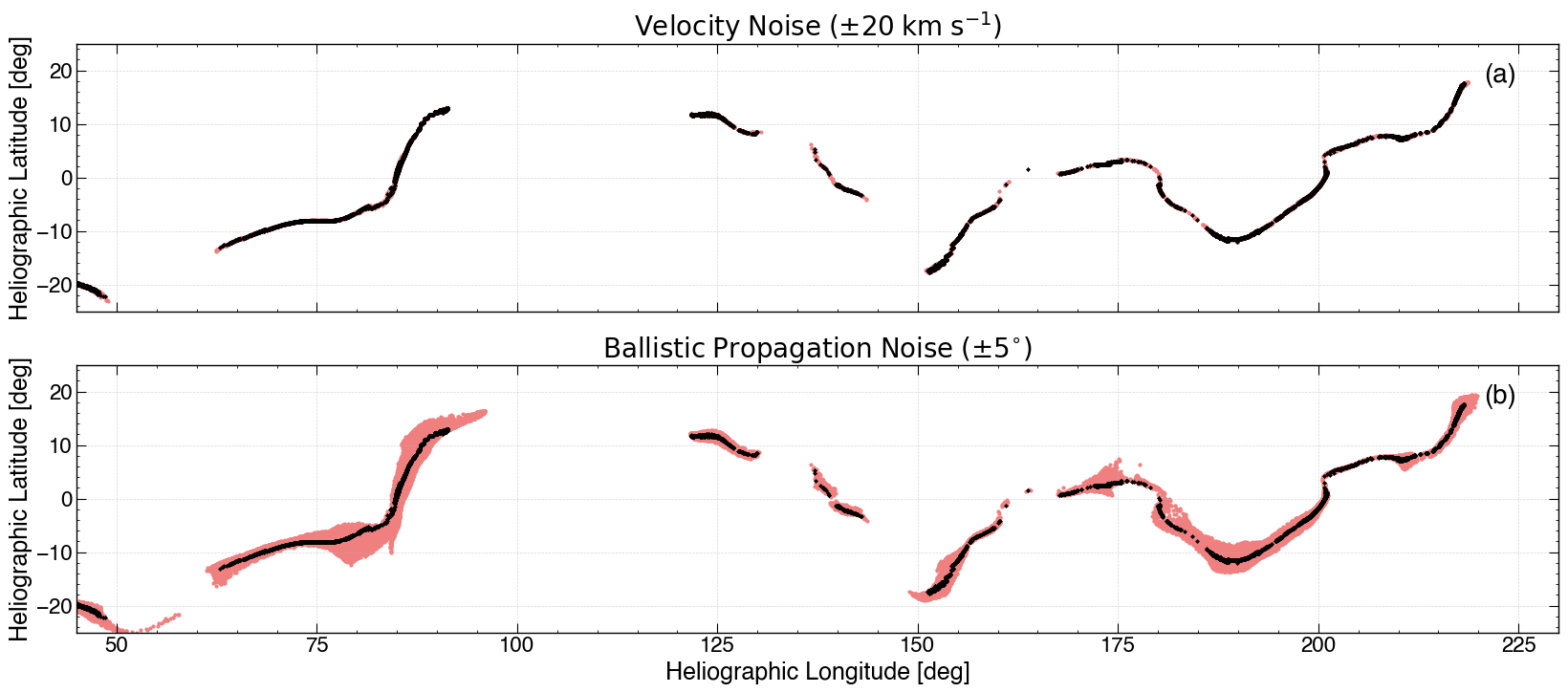}
  \caption{Comparison of the estimated PFSS footpoints used for this study (black) with footpoints estimated with different induced errors (shown in pink). \textit{Panel (a):} PFSS footpoint estimations with $\pm 20$ {\kms} of random noise induced in the PSP radial velocity measurement used for ballistic propagation. \textit{Panel (b):} PFSS footpoint estimations with $\pm 5^{\circ}$ of random noise induced in the source surface location based on ballistic propagation ballistic propagation.}
  \label{fig: footpoint-error}
\end{figure*}

In Figure~\ref{fig: footpoint-error}, we compare estimated footpoints from the PFSS modeling method with footpoints produced with some random noise. We compare footpoints produced with noise both in the PSP velocity measurement used for the ballistic propagation model, along with errors in footpoints from adding $\pm 5^{\circ}$ in variation of the longitude and latitude source surface points. Similar to Figure~\ref{fig: mhd-comp}, we see the jump in footpoints at $\sim$ 165{\degree} characteristic of a streamer crossing, and that these errors do not impact the resulting source region of the observed solar wind at PSP and Solar Orbiter.

\bibliography{ms}{}

\begin{thebibliography}{}
\expandafter\ifx\csname natexlab\endcsname\relax\def\natexlab#1{#1}\fi
\providecommand{\url}[1]{\href{#1}{#1}}
\providecommand{\dodoi}[1]{doi:~\href{http://doi.org/#1}{\nolinkurl{#1}}}
\providecommand{\doeprint}[1]{\href{http://ascl.net/#1}{\nolinkurl{http://ascl.net/#1}}}
\providecommand{\doarXiv}[1]{\href{https://arxiv.org/abs/#1}{\nolinkurl{https://arxiv.org/abs/#1}}}

\bibitem[{{Abbo} {et~al.}(2016){Abbo}, {Ofman}, {Antiochos}, {Hansteen}, {Harra}, {Ko}, {Lapenta}, {Li}, {Riley}, {Strachan}, {von Steiger}, \& {Wang}}]{Abbo-2016}
{Abbo}, L., {Ofman}, L., {Antiochos}, S.~K., {et~al.} 2016, \ssr, 201, 55, \dodoi{10.1007/s11214-016-0264-1}

\bibitem[{Alterman \& Kasper(2019)}]{Alterman-2019}
Alterman, B.~L., \& Kasper, J.~C. 2019, The Astrophysical Journal Letters, 879, L6, \dodoi{10.3847/2041-8213/ab2391}

\bibitem[{{Alterman} {et~al.}(2018){Alterman}, {Kasper}, {Stevens}, \& {Koval}}]{Alterman-2018}
{Alterman}, B.~L., {Kasper}, J.~C., {Stevens}, M.~L., \& {Koval}, A. 2018, \apj, 864, 112, \dodoi{10.3847/1538-4357/aad23f}

\bibitem[{{Altschuler} \& {Newkirk}(1969)}]{Altschuler-1969}
{Altschuler}, M.~D., \& {Newkirk}, G. 1969, \solphys, 9, 131, \dodoi{10.1007/BF00145734}

\bibitem[{{Annex} {et~al.}(2020){Annex}, {Pearson}, {Seignovert}, {Carcich}, {Eichhorn}, {Mapel}, {von Forstner}, {McAuliffe}, {del Rio}, {Berry}, {Aye}, {Stefko}, {de Val-Borro}, {Kulumani}, \& {Murakami}}]{spiceypy}
{Annex}, A., {Pearson}, B., {Seignovert}, B., {et~al.} 2020, The Journal of Open Source Software, 5, 2050, \dodoi{10.21105/joss.02050}

\bibitem[{{Antiochos} {et~al.}(2012){Antiochos}, {Linker}, {Lionello}, {Miki{\'c}}, {Titov}, \& {Zurbuchen}}]{Antiochos-2012}
{Antiochos}, S.~K., {Linker}, J.~A., {Lionello}, R., {et~al.} 2012, \ssr, 172, 169, \dodoi{10.1007/s11214-011-9795-7}

\bibitem[{{Arge} {et~al.}(2013){Arge}, {Henney}, {Hernandez}, {Toussaint}, {Koller}, \& {Godinez}}]{Arge-2013}
{Arge}, C.~N., {Henney}, C.~J., {Hernandez}, I.~G., {et~al.} 2013, in American Institute of Physics Conference Series, Vol. 1539, Solar Wind 13, ed. G.~P. {Zank}, J.~{Borovsky}, R.~{Bruno}, J.~{Cirtain}, S.~{Cranmer}, H.~{Elliott}, J.~{Giacalone}, W.~{Gonzalez}, G.~{Li}, E.~{Marsch}, E.~{Moebius}, N.~{Pogorelov}, J.~{Spann}, \& O.~{Verkhoglyadova}, 11--14, \dodoi{10.1063/1.4810977}

\bibitem[{{Arge} {et~al.}(2010){Arge}, {Henney}, {Koller}, {Compeau}, {Young}, {MacKenzie}, {Fay}, \& {Harvey}}]{Arge-2010}
{Arge}, C.~N., {Henney}, C.~J., {Koller}, J., {et~al.} 2010, in American Institute of Physics Conference Series, Vol. 1216, Twelfth International Solar Wind Conference, ed. M.~{Maksimovic}, K.~{Issautier}, N.~{Meyer-Vernet}, M.~{Moncuquet}, \& F.~{Pantellini}, 343--346, \dodoi{10.1063/1.3395870}

\bibitem[{{Arge} {et~al.}(2011){Arge}, {Henney}, {Koller}, {Toussaint}, {Harvey}, \& {Young}}]{Arge-2011}
{Arge}, C.~N., {Henney}, C.~J., {Koller}, J., {et~al.} 2011, in Astronomical Society of the Pacific Conference Series, Vol. 444, 5th International Conference of Numerical Modeling of Space Plasma Flows (ASTRONUM 2010), ed. N.~V. {Pogorelov}, E.~{Audit}, \& G.~P. {Zank}, 99

\bibitem[{{Arge} {et~al.}(2004){Arge}, {Luhmann}, {Odstrcil}, {Schrijver}, \& {Li}}]{Arge-2004}
{Arge}, C.~N., {Luhmann}, J.~G., {Odstrcil}, D., {Schrijver}, C.~J., \& {Li}, Y. 2004, Journal of Atmospheric and Solar-Terrestrial Physics, 66, 1295, \dodoi{10.1016/j.jastp.2004.03.018}

\bibitem[{Arge {et~al.}(2003)Arge, Odstrcil, Pizzo, \& Mayer}]{Arge-2003}
Arge, C.~N., Odstrcil, D., Pizzo, V.~J., \& Mayer, L.~R. 2003, AIP Conference Proceedings, 679, 190, \dodoi{10.1063/1.1618574}

\bibitem[{{Arge} \& {Pizzo}(2000)}]{Arge-2000}
{Arge}, C.~N., \& {Pizzo}, V.~J. 2000, \jgr, 105, 10465, \dodoi{10.1029/1999JA000262}

\bibitem[{{Asplund} {et~al.}(2021){Asplund}, {Amarsi}, \& {Grevesse}}]{Asplund-2021}
{Asplund}, M., {Amarsi}, A.~M., \& {Grevesse}, N. 2021, \aap, 653, A141, \dodoi{10.1051/0004-6361/202140445}

\bibitem[{{Asplund} {et~al.}(2009){Asplund}, {Grevesse}, {Sauval}, \& {Scott}}]{Asplund-2009}
{Asplund}, M., {Grevesse}, N., {Sauval}, A.~J., \& {Scott}, P. 2009, \araa, 47, 481, \dodoi{10.1146/annurev.astro.46.060407.145222}

\bibitem[{{Astropy Collaboration} {et~al.}(2013){Astropy Collaboration}, {Robitaille}, {Tollerud}, {Greenfield}, {Droettboom}, {Bray}, {Aldcroft}, {Davis}, {Ginsburg}, {Price-Whelan}, {Kerzendorf}, {Conley}, {Crighton}, {Barbary}, {Muna}, {Ferguson}, {Grollier}, {Parikh}, {Nair}, {Unther}, {Deil}, {Woillez}, {Conseil}, {Kramer}, {Turner}, {Singer}, {Fox}, {Weaver}, {Zabalza}, {Edwards}, {Azalee Bostroem}, {Burke}, {Casey}, {Crawford}, {Dencheva}, {Ely}, {Jenness}, {Labrie}, {Lim}, {Pierfederici}, {Pontzen}, {Ptak}, {Refsdal}, {Servillat}, \& {Streicher}}]{astropy:2013}
{Astropy Collaboration}, {Robitaille}, T.~P., {Tollerud}, E.~J., {et~al.} 2013, \aap, 558, A33, \dodoi{10.1051/0004-6361/201322068}

\bibitem[{{Astropy Collaboration} {et~al.}(2018){Astropy Collaboration}, {Price-Whelan}, {Sip{\H{o}}cz}, {G{\"u}nther}, {Lim}, {Crawford}, {Conseil}, {Shupe}, {Craig}, {Dencheva}, {Ginsburg}, {VanderPlas}, {Bradley}, {P{\'e}rez-Su{\'a}rez}, {de Val-Borro}, {Aldcroft}, {Cruz}, {Robitaille}, {Tollerud}, {Ardelean}, {Babej}, {Bach}, {Bachetti}, {Bakanov}, {Bamford}, {Barentsen}, {Barmby}, {Baumbach}, {Berry}, {Biscani}, {Boquien}, {Bostroem}, {Bouma}, {Brammer}, {Bray}, {Breytenbach}, {Buddelmeijer}, {Burke}, {Calderone}, {Cano Rodr{\'\i}guez}, {Cara}, {Cardoso}, {Cheedella}, {Copin}, {Corrales}, {Crichton}, {D'Avella}, {Deil}, {Depagne}, {Dietrich}, {Donath}, {Droettboom}, {Earl}, {Erben}, {Fabbro}, {Ferreira}, {Finethy}, {Fox}, {Garrison}, {Gibbons}, {Goldstein}, {Gommers}, {Greco}, {Greenfield}, {Groener}, {Grollier}, {Hagen}, {Hirst}, {Homeier}, {Horton}, {Hosseinzadeh}, {Hu}, {Hunkeler}, {Ivezi{\'c}}, {Jain}, {Jenness}, {Kanarek}, {Kendrew}, {Kern}, {Kerzendorf}, {Khvalko}, {King}, {Kirkby}, {Kulkarni},
  {Kumar}, {Lee}, {Lenz}, {Littlefair}, {Ma}, {Macleod}, {Mastropietro}, {McCully}, {Montagnac}, {Morris}, {Mueller}, {Mumford}, {Muna}, {Murphy}, {Nelson}, {Nguyen}, {Ninan}, {N{\"o}the}, {Ogaz}, {Oh}, {Parejko}, {Parley}, {Pascual}, {Patil}, {Patil}, {Plunkett}, {Prochaska}, {Rastogi}, {Reddy Janga}, {Sabater}, {Sakurikar}, {Seifert}, {Sherbert}, {Sherwood-Taylor}, {Shih}, {Sick}, {Silbiger}, {Singanamalla}, {Singer}, {Sladen}, {Sooley}, {Sornarajah}, {Streicher}, {Teuben}, {Thomas}, {Tremblay}, {Turner}, {Terr{\'o}n}, {van Kerkwijk}, {de la Vega}, {Watkins}, {Weaver}, {Whitmore}, {Woillez}, {Zabalza}, \& {Astropy Contributors}}]{astropy:2018}
{Astropy Collaboration}, {Price-Whelan}, A.~M., {Sip{\H{o}}cz}, B.~M., {et~al.} 2018, \aj, 156, 123, \dodoi{10.3847/1538-3881/aabc4f}

\bibitem[{{Astropy Collaboration} {et~al.}(2022){Astropy Collaboration}, {Price-Whelan}, {Lim}, {Earl}, {Starkman}, {Bradley}, {Shupe}, {Patil}, {Corrales}, {Brasseur}, {N{\"o}the}, {Donath}, {Tollerud}, {Morris}, {Ginsburg}, {Vaher}, {Weaver}, {Tocknell}, {Jamieson}, {van Kerkwijk}, {Robitaille}, {Merry}, {Bachetti}, {G{\"u}nther}, {Aldcroft}, {Alvarado-Montes}, {Archibald}, {B{\'o}di}, {Bapat}, {Barentsen}, {Baz{\'a}n}, {Biswas}, {Boquien}, {Burke}, {Cara}, {Cara}, {Conroy}, {Conseil}, {Craig}, {Cross}, {Cruz}, {D'Eugenio}, {Dencheva}, {Devillepoix}, {Dietrich}, {Eigenbrot}, {Erben}, {Ferreira}, {Foreman-Mackey}, {Fox}, {Freij}, {Garg}, {Geda}, {Glattly}, {Gondhalekar}, {Gordon}, {Grant}, {Greenfield}, {Groener}, {Guest}, {Gurovich}, {Handberg}, {Hart}, {Hatfield-Dodds}, {Homeier}, {Hosseinzadeh}, {Jenness}, {Jones}, {Joseph}, {Kalmbach}, {Karamehmetoglu}, {Ka{\l}uszy{\'n}ski}, {Kelley}, {Kern}, {Kerzendorf}, {Koch}, {Kulumani}, {Lee}, {Ly}, {Ma}, {MacBride}, {Maljaars}, {Muna}, {Murphy}, {Norman},
  {O'Steen}, {Oman}, {Pacifici}, {Pascual}, {Pascual-Granado}, {Patil}, {Perren}, {Pickering}, {Rastogi}, {Roulston}, {Ryan}, {Rykoff}, {Sabater}, {Sakurikar}, {Salgado}, {Sanghi}, {Saunders}, {Savchenko}, {Schwardt}, {Seifert-Eckert}, {Shih}, {Jain}, {Shukla}, {Sick}, {Simpson}, {Singanamalla}, {Singer}, {Singhal}, {Sinha}, {Sip{\H{o}}cz}, {Spitler}, {Stansby}, {Streicher}, {{\v{S}}umak}, {Swinbank}, {Taranu}, {Tewary}, {Tremblay}, {de Val-Borro}, {Van Kooten}, {Vasovi{\'c}}, {Verma}, {de Miranda Cardoso}, {Williams}, {Wilson}, {Winkel}, {Wood-Vasey}, {Xue}, {Yoachim}, {Zhang}, {Zonca}, \& {Astropy Project Contributors}}]{astropy:2022}
{Astropy Collaboration}, {Price-Whelan}, A.~M., {Lim}, P.~L., {et~al.} 2022, \apj, 935, 167, \dodoi{10.3847/1538-4357/ac7c74}

\bibitem[{{Badman} {et~al.}(2020){Badman}, {Bale}, {Mart{\'\i}nez Oliveros}, {Panasenco}, {Velli}, {Stansby}, {Buitrago-Casas}, {R{\'e}ville}, {Bonnell}, {Case}, {Dudok de Wit}, {Goetz}, {Harvey}, {Kasper}, {Korreck}, {Larson}, {Livi}, {MacDowall}, {Malaspina}, {Pulupa}, {Stevens}, \& {Whittlesey}}]{Badman-2020}
{Badman}, S.~T., {Bale}, S.~D., {Mart{\'\i}nez Oliveros}, J.~C., {et~al.} 2020, \apjs, 246, 23, \dodoi{10.3847/1538-4365/ab4da7}

\bibitem[{{Badman} {et~al.}(2022){Badman}, {Brooks}, {Poirier}, {Warren}, {Petrie}, {Rouillard}, {Nick Arge}, {Bale}, {de Pablos Ag{\"u}ero}, {Harra}, {Jones}, {Kouloumvakos}, {Riley}, {Panasenco}, {Velli}, \& {Wallace}}]{Badman-2022}
{Badman}, S.~T., {Brooks}, D.~H., {Poirier}, N., {et~al.} 2022, \apj, 932, 135, \dodoi{10.3847/1538-4357/ac6610}

\bibitem[{{Badman} {et~al.}(2023){Badman}, {Riley}, {Jones}, {Kim}, {Allen}, {Arge}, {Bale}, {Henney}, {Kasper}, {Mostafavi}, {Pogorelov}, {Raouafi}, {Stevens}, \& {Verniero}}]{Badman-2023}
{Badman}, S.~T., {Riley}, P., {Jones}, S.~I., {et~al.} 2023, Journal of Geophysical Research (Space Physics), 128, e2023JA031359, \dodoi{10.1029/2023JA031359}

\bibitem[{{Bale} {et~al.}(2016){Bale}, {Goetz}, {Harvey}, {Turin}, {Bonnell}, {Dudok de Wit}, {Ergun}, {MacDowall}, {Pulupa}, {Andre}, {Bolton}, {Bougeret}, {Bowen}, {Burgess}, {Cattell}, {Chandran}, {Chaston}, {Chen}, {Choi}, {Connerney}, {Cranmer}, {Diaz-Aguado}, {Donakowski}, {Drake}, {Farrell}, {Fergeau}, {Fermin}, {Fischer}, {Fox}, {Glaser}, {Goldstein}, {Gordon}, {Hanson}, {Harris}, {Hayes}, {Hinze}, {Hollweg}, {Horbury}, {Howard}, {Hoxie}, {Jannet}, {Karlsson}, {Kasper}, {Kellogg}, {Kien}, {Klimchuk}, {Krasnoselskikh}, {Krucker}, {Lynch}, {Maksimovic}, {Malaspina}, {Marker}, {Martin}, {Martinez-Oliveros}, {McCauley}, {McComas}, {McDonald}, {Meyer-Vernet}, {Moncuquet}, {Monson}, {Mozer}, {Murphy}, {Odom}, {Oliverson}, {Olson}, {Parker}, {Pankow}, {Phan}, {Quataert}, {Quinn}, {Ruplin}, {Salem}, {Seitz}, {Sheppard}, {Siy}, {Stevens}, {Summers}, {Szabo}, {Timofeeva}, {Vaivads}, {Velli}, {Yehle}, {Werthimer}, \& {Wygant}}]{Bale-2016}
{Bale}, S.~D., {Goetz}, K., {Harvey}, P.~R., {et~al.} 2016, \ssr, 204, 49, \dodoi{10.1007/s11214-016-0244-5}

\bibitem[{{Bame} {et~al.}(1977){Bame}, {Asbridge}, {Feldman}, \& {Gosling}}]{Bame-1997}
{Bame}, S.~J., {Asbridge}, J.~R., {Feldman}, W.~C., \& {Gosling}, J.~T. 1977, \jgr, 82, 1487, \dodoi{10.1029/JA082i010p01487}

\bibitem[{{Bavassano} {et~al.}(1998){Bavassano}, {Pietropaolo}, \& {Bruno}}]{Bavassano-1998}
{Bavassano}, B., {Pietropaolo}, E., \& {Bruno}, R. 1998, \jgr, 103, 6521, \dodoi{10.1029/97JA03029}

\bibitem[{{Bavassano} {et~al.}(1997){Bavassano}, {Woo}, \& {Bruno}}]{Bavassano-1997}
{Bavassano}, B., {Woo}, R., \& {Bruno}, R. 1997, \grl, 24, 1655, \dodoi{10.1029/97GL01630}

\bibitem[{{Bochsler}(2007)}]{Bochsler-2007}
{Bochsler}, P. 2007, \aap, 471, 315, \dodoi{10.1051/0004-6361:20077772}

\bibitem[{{Bochsler} {et~al.}(1986){Bochsler}, {Geiss}, \& {Kunz}}]{Bochsler-1986}
{Bochsler}, P., {Geiss}, J., \& {Kunz}, S. 1986, \solphys, 103, 177, \dodoi{10.1007/BF00154867}

\bibitem[{Borovsky(2012)}]{Borovsky-2012}
Borovsky, J.~E. 2012, Journal of Geophysical Research: Space Physics, 117, \dodoi{https://doi.org/10.1029/2012JA017525}

\bibitem[{{Borrini} {et~al.}(1981){Borrini}, {Gosling}, {Bame}, {Feldman}, \& {Wilcox}}]{Borrini-1981}
{Borrini}, G., {Gosling}, J.~T., {Bame}, S.~J., {Feldman}, W.~C., \& {Wilcox}, J.~M. 1981, \jgr, 86, 4565, \dodoi{10.1029/JA086iA06p04565}

\bibitem[{{Bruno} {et~al.}(1986){Bruno}, {Villante}, {Bavassano}, {Schwenn}, \& {Mariani}}]{Bruno-1986}
{Bruno}, R., {Villante}, U., {Bavassano}, B., {Schwenn}, R., \& {Mariani}, F. 1986, \solphys, 104, 431, \dodoi{10.1007/BF00159093}

\bibitem[{{Buergi} \& {Geiss}(1986)}]{Buergi-1986}
{Buergi}, A., \& {Geiss}, J. 1986, \solphys, 103, 347, \dodoi{10.1007/BF00147835}

\bibitem[{Chen {et~al.}(2013)Chen, Bale, Salem, \& Maruca}]{Chen-2013}
Chen, C. H.~K., Bale, S.~D., Salem, C.~S., \& Maruca, B.~A. 2013, The Astrophysical Journal, 770, 125, \dodoi{10.1088/0004-637X/770/2/125}

\bibitem[{{Chen} {et~al.}(2021){Chen}, {Chandran}, {Woodham}, {Jones}, {Perez}, {Bourouaine}, {Bowen}, {Klein}, {Moncuquet}, {Kasper}, \& {Bale}}]{Chen-2021}
{Chen}, C.~H.~K., {Chandran}, B.~D.~G., {Woodham}, L.~D., {et~al.} 2021, \aap, 650, L3, \dodoi{10.1051/0004-6361/202039872}

\bibitem[{{Chen} {et~al.}(2003){Chen}, {Esser}, \& {Hu}}]{Chen-2003}
{Chen}, Y., {Esser}, R., \& {Hu}, Y. 2003, \apj, 582, 467, \dodoi{10.1086/344642}

\bibitem[{Community {et~al.}(2020)Community, Barnes, Bobra, Christe, Freij, Hayes, Ireland, Mumford, Perez-Suarez, Ryan, Shih, Contributors), Chanda, Glogowski, Hewett, Hughitt, Hill, Hiware, Inglis, Kirk, Konge, Mason, Maloney, Murray, Panda, Park, Pereira, Reardon, Savage, Sip{\H o}cz, Stansby, Jain, Taylor, Yadav, Rajul, Dang, \& Contributors)}]{sunpy}
Community, T.~S., Barnes, W.~T., Bobra, M.~G., {et~al.} 2020, The Astrophysical Journal, 890, 68, \dodoi{10.3847/1538-4357/ab4f7a}

\bibitem[{Cranmer \& Winebarger(2019)}]{Cranmer-2019}
Cranmer, S.~R., \& Winebarger, A.~R. 2019, Annual Review of Astronomy and Astrophysics, 57, 157, \dodoi{10.1146/annurev-astro-091918-104416}

\bibitem[{{Crooker} {et~al.}(2012){Crooker}, {Antiochos}, {Zhao}, \& {Neugebauer}}]{Crooker-2012}
{Crooker}, N.~U., {Antiochos}, S.~K., {Zhao}, X., \& {Neugebauer}, M. 2012, Journal of Geophysical Research (Space Physics), 117, A04104, \dodoi{10.1029/2011JA017236}

\bibitem[{{Culhane} {et~al.}(2014){Culhane}, {Brooks}, {van Driel-Gesztelyi}, {D{\'e}moulin}, {Baker}, {DeRosa}, {Mandrini}, {Zhao}, \& {Zurbuchen}}]{Culhane-2014}
{Culhane}, J.~L., {Brooks}, D.~H., {van Driel-Gesztelyi}, L., {et~al.} 2014, \solphys, 289, 3799, \dodoi{10.1007/s11207-014-0551-5}

\bibitem[{{D'Amicis} \& {Bruno}(2015)}]{DAmicis-2015}
{D'Amicis}, R., \& {Bruno}, R. 2015, \apj, 805, 84, \dodoi{10.1088/0004-637X/805/1/84}

\bibitem[{D'Amicis {et~al.}(2018)D'Amicis, Matteini, \& Bruno}]{DAmicis-2018}
D'Amicis, R., Matteini, L., \& Bruno, R. 2018, Monthly Notices of the Royal Astronomical Society, 483, 4665, \dodoi{10.1093/mnras/sty3329}

\bibitem[{D'Amicis {et~al.}(2021)D'Amicis, Perrone, Bruno, \& Velli}]{DAmicis-2021}
D'Amicis, R., Perrone, D., Bruno, R., \& Velli, M. 2021, Journal of Geophysical Research: Space Physics, 126, e2020JA028996, \dodoi{https://doi.org/10.1029/2020JA028996}

\bibitem[{{D'Amicis} {et~al.}(2021){D'Amicis}, {Bruno}, {Panasenco}, {Telloni}, {Perrone}, {Marcucci}, {Woodham}, {Velli}, {De Marco}, {Jagarlamudi}, {Coco}, {Owen}, {Louarn}, {Livi}, {Horbury}, {Andr{\'e}}, {Angelini}, {Evans}, {Fedorov}, {Genot}, {Lavraud}, {Matteini}, {M{\"u}ller}, {O'Brien}, {Pezzi}, {Rouillard}, {Sorriso-Valvo}, {Tenerani}, {Verscharen}, \& {Zouganelis}}]{DAmicis-2021aa}
{D'Amicis}, R., {Bruno}, R., {Panasenco}, O., {et~al.} 2021, \aap, 656, A21, \dodoi{10.1051/0004-6361/202140938}

\bibitem[{{Elsasser}(1950)}]{Elsasser-1950}
{Elsasser}, W.~M. 1950, Physical Review, 79, 183, \dodoi{10.1103/PhysRev.79.183}

\bibitem[{{Feldman}(1992)}]{Feldman-1992}
{Feldman}, U. 1992, \physscr, 46, 202, \dodoi{10.1088/0031-8949/46/3/002}

\bibitem[{{Feldman} \& {Laming}(2000)}]{Feldman-2000}
{Feldman}, U., \& {Laming}, J.~M. 2000, \physscr, 61, 222, \dodoi{10.1238/Physica.Regular.061a00222}

\bibitem[{{Feldman} \& {Marsch}(1997)}]{Feldman-1997}
{Feldman}, W.~C., \& {Marsch}, E. 1997, in Cosmic Winds and the Heliosphere, 617

\bibitem[{{Fox} {et~al.}(2016){Fox}, {Velli}, {Bale}, {Decker}, {Driesman}, {Howard}, {Kasper}, {Kinnison}, {Kusterer}, {Lario}, {Lockwood}, {McComas}, {Raouafi}, \& {Szabo}}]{Fox-2016}
{Fox}, N.~J., {Velli}, M.~C., {Bale}, S.~D., {et~al.} 2016, \ssr, 204, 7, \dodoi{10.1007/s11214-015-0211-6}

\bibitem[{Garton {et~al.}(2018)Garton, Murray, \& Gallagher}]{Garton-2018}
Garton, T.~M., Murray, S.~A., \& Gallagher, P.~T. 2018, The Astrophysical Journal Letters, 869, L12, \dodoi{10.3847/2041-8213/aaf39a}

\bibitem[{{Gary} {et~al.}(2000){Gary}, {Yin}, {Winske}, \& {Reisenfeld}}]{Gary-2000}
{Gary}, S.~P., {Yin}, L., {Winske}, D., \& {Reisenfeld}, D.~B. 2000, \grl, 27, 1355, \dodoi{10.1029/2000GL000019}

\bibitem[{Geiss {et~al.}(1995)Geiss, Gloeckler, \& Von~Steiger}]{Geiss-1995}
Geiss, J., Gloeckler, G., \& Von~Steiger, R. 1995, Space Science Reviews, 72, 49, \dodoi{10.1007/BF00768753}

\bibitem[{{Gloeckler} \& {Geiss}(1989)}]{Gloeckler-1989}
{Gloeckler}, G., \& {Geiss}, J. 1989, in American Institute of Physics Conference Series, Vol. 183, Cosmic Abundances of Matter, ed. C.~J. {Waddington}, 49--71, \dodoi{10.1063/1.37985}

\bibitem[{{Gosling} {et~al.}(1981){Gosling}, {Borrini}, {Asbridge}, {Bame}, {Feldman}, \& {Hansen}}]{Gosling-1981}
{Gosling}, J.~T., {Borrini}, G., {Asbridge}, J.~R., {et~al.} 1981, \jgr, 86, 5438, \dodoi{10.1029/JA086iA07p05438}

\bibitem[{Harris {et~al.}(2020)Harris, Millman, van~der Walt, Gommers, Virtanen, Cournapeau, Wieser, Taylor, Berg, Smith, Kern, Picus, Hoyer, van Kerkwijk, Brett, Haldane, del R{\'\i}o, Wiebe, Peterson, G{\'e}rard-Marchant, Sheppard, Reddy, Weckesser, Abbasi, Gohlke, \& Oliphant}]{numpy}
Harris, C.~R., Millman, K.~J., van~der Walt, S.~J., {et~al.} 2020, Nature, 585, 357, \dodoi{10.1038/s41586-020-2649-2}

\bibitem[{{Harvey} {et~al.}(1996){Harvey}, {Hill}, {Hubbard}, {Kennedy}, {Leibacher}, {Pintar}, {Gilman}, {Noyes}, {Title}, {Toomre}, {Ulrich}, {Bhatnagar}, {Kennewell}, {Marquette}, {Patron}, {Saa}, \& {Yasukawa}}]{Harvey-1996}
{Harvey}, J.~W., {Hill}, F., {Hubbard}, R.~P., {et~al.} 1996, Science, 272, 1284, \dodoi{10.1126/science.272.5266.1284}

\bibitem[{Hickmann {et~al.}(2015)Hickmann, Godinez, Henney, \& Arge}]{Hickmann-2015}
Hickmann, K.~S., Godinez, H.~C., Henney, C.~J., \& Arge, C.~N. 2015, Solar Physics, 290, 1105, \dodoi{10.1007/s11207-015-0666-3}

\bibitem[{{Hoeksema}(1984)}]{Hoeksema-1984}
{Hoeksema}, J.~T. 1984, PhD thesis, Stanford University, California

\bibitem[{{Horbury} {et~al.}(2020){Horbury}, {O'Brien}, {Carrasco Blazquez}, {Bendyk}, {Brown}, {Hudson}, {Evans}, {Oddy}, {Carr}, {Beek}, {Cupido}, {Bhattacharya}, {Dominguez}, {Matthews}, {Myklebust}, {Whiteside}, {Bale}, {Baumjohann}, {Burgess}, {Carbone}, {Cargill}, {Eastwood}, {Erd{\"o}s}, {Fletcher}, {Forsyth}, {Giacalone}, {Glassmeier}, {Goldstein}, {Hoeksema}, {Lockwood}, {Magnes}, {Maksimovic}, {Marsch}, {Matthaeus}, {Murphy}, {Nakariakov}, {Owen}, {Owens}, {Rodriguez-Pacheco}, {Richter}, {Riley}, {Russell}, {Schwartz}, {Vainio}, {Velli}, {Vennerstrom}, {Walsh}, {Wimmer-Schweingruber}, {Zank}, {M{\"u}ller}, {Zouganelis}, \& {Walsh}}]{Horbury-2020}
{Horbury}, T.~S., {O'Brien}, H., {Carrasco Blazquez}, I., {et~al.} 2020, \aap, 642, A9, \dodoi{10.1051/0004-6361/201937257}

\bibitem[{{Howard} {et~al.}(2008){Howard}, {Moses}, {Vourlidas}, {Newmark}, {Socker}, {Plunkett}, {Korendyke}, {Cook}, {Hurley}, {Davila}, {Thompson}, {St Cyr}, {Mentzell}, {Mehalick}, {Lemen}, {Wuelser}, {Duncan}, {Tarbell}, {Wolfson}, {Moore}, {Harrison}, {Waltham}, {Lang}, {Davis}, {Eyles}, {Mapson-Menard}, {Simnett}, {Halain}, {Defise}, {Mazy}, {Rochus}, {Mercier}, {Ravet}, {Delmotte}, {Auchere}, {Delaboudiniere}, {Bothmer}, {Deutsch}, {Wang}, {Rich}, {Cooper}, {Stephens}, {Maahs}, {Baugh}, {McMullin}, \& {Carter}}]{Howard-2008}
{Howard}, R.~A., {Moses}, J.~D., {Vourlidas}, A., {et~al.} 2008, \ssr, 136, 67, \dodoi{10.1007/s11214-008-9341-4}

\bibitem[{Hundhausen(1968)}]{Hundhausen-1968}
Hundhausen, A.~J. 1968, Space Science Reviews, 8, 690, \dodoi{10.1007/BF00175116}

\bibitem[{Hunter(2007)}]{mpl}
Hunter, J.~D. 2007, Computing in Science \& Engineering, 9, 90, \dodoi{10.1109/MCSE.2007.55}

\bibitem[{Kasper {et~al.}(2012)Kasper, Stevens, Korreck, Maruca, Kiefer, Schwadron, \& Lepri}]{Kasper-2012}
Kasper, J.~C., Stevens, M.~L., Korreck, K.~E., {et~al.} 2012, The Astrophysical Journal, 745, 162, \dodoi{10.1088/0004-637X/745/2/162}

\bibitem[{{Kasper} {et~al.}(2007){Kasper}, {Stevens}, {Lazarus}, {Steinberg}, \& {Ogilvie}}]{Kasper-2007}
{Kasper}, J.~C., {Stevens}, M.~L., {Lazarus}, A.~J., {Steinberg}, J.~T., \& {Ogilvie}, K.~W. 2007, \apj, 660, 901, \dodoi{10.1086/510842}

\bibitem[{Kasper {et~al.}(2016)Kasper, Abiad, Austin, Balat-Pichelin, Bale, Belcher, Berg, Bergner, Berthomier, Bookbinder, Brodu, Caldwell, Case, Chandran, Cheimets, Cirtain, Cranmer, Curtis, Daigneau, Dalton, Dasgupta, DeTomaso, Diaz-Aguado, Djordjevic, Donaskowski, Effinger, Florinski, Fox, Freeman, Gallagher, Gary, Gauron, Gates, Goldstein, Golub, Gordon, Gurnee, Guth, Halekas, Hatch, Heerikuisen, Ho, Hu, Johnson, Jordan, Korreck, Larson, Lazarus, Li, Livi, Ludlam, Maksimovic, McFadden, Marchant, Maruca, McComas, Messina, Mercer, Park, Peddie, Pogorelov, Reinhart, Richardson, Robinson, Rosen, Skoug, Slagle, Steinberg, Stevens, Szabo, Taylor, Tiu, Turin, Velli, Webb, Whittlesey, Wright, Wu, \& Zank}]{Kasper-2016}
Kasper, J.~C., Abiad, R., Austin, G., {et~al.} 2016, Space Science Reviews, 204, 131, \dodoi{10.1007/s11214-015-0206-3}

\bibitem[{{Kasper} {et~al.}(2019){Kasper}, {Bale}, {Belcher}, {Berthomier}, {Case}, {Chandran}, {Curtis}, {Gallagher}, {Gary}, {Golub}, {Halekas}, {Ho}, {Horbury}, {Hu}, {Huang}, {Klein}, {Korreck}, {Larson}, {Livi}, {Maruca}, {Lavraud}, {Louarn}, {Maksimovic}, {Martinovic}, {McGinnis}, {Pogorelov}, {Richardson}, {Skoug}, {Steinberg}, {Stevens}, {Szabo}, {Velli}, {Whittlesey}, {Wright}, {Zank}, {MacDowall}, {McComas}, {McNutt}, {Pulupa}, {Raouafi}, \& {Schwadron}}]{Kasper-2019}
{Kasper}, J.~C., {Bale}, S.~D., {Belcher}, J.~W., {et~al.} 2019, \nat, 576, 228, \dodoi{10.1038/s41586-019-1813-z}

\bibitem[{{Ko} {et~al.}(2014){Ko}, {Muglach}, {Wang}, {Young}, \& {Lepri}}]{Ko-2014}
{Ko}, Y.-K., {Muglach}, K., {Wang}, Y.-M., {Young}, P.~R., \& {Lepri}, S.~T. 2014, \apj, 787, 121, \dodoi{10.1088/0004-637X/787/2/121}

\bibitem[{{Ko} {et~al.}(2006){Ko}, {Raymond}, {Zurbuchen}, {Riley}, {Raines}, \& {Strachan}}]{Ko-2006}
{Ko}, Y.-K., {Raymond}, J.~C., {Zurbuchen}, T.~H., {et~al.} 2006, \apj, 646, 1275, \dodoi{10.1086/505021}

\bibitem[{{Koukras} {et~al.}(2022){Koukras}, {Dolla}, \& {Keppens}}]{Koukras-2022}
{Koukras}, A., {Dolla}, L., \& {Keppens}, R. 2022, in SHINE 2022 Workshop, 68, \dodoi{10.48550/arXiv.2212.11553}

\bibitem[{Laming(2015)}]{Laming-2015}
Laming, J.~M. 2015, Living Reviews in Solar Physics, 12, 2, \dodoi{10.1007/lrsp-2015-2}

\bibitem[{{Laming}(2017)}]{Laming-2017}
{Laming}, J.~M. 2017, \apj, 844, 153, \dodoi{10.3847/1538-4357/aa7cf1}

\bibitem[{{Laming} {et~al.}(2019){Laming}, {Vourlidas}, {Korendyke}, {Chua}, {Cranmer}, {Ko}, {Kuroda}, {Provornikova}, {Raymond}, {Raouafi}, {Strachan}, {Tun-Beltran}, {Weberg}, \& {Wood}}]{Laming-2019}
{Laming}, J.~M., {Vourlidas}, A., {Korendyke}, C., {et~al.} 2019, \apj, 879, 124, \dodoi{10.3847/1538-4357/ab23f1}

\bibitem[{{Landi} {et~al.}(2012){Landi}, {Gruesbeck}, {Lepri}, {Zurbuchen}, \& {Fisk}}]{Landi-2012b}
{Landi}, E., {Gruesbeck}, J.~R., {Lepri}, S.~T., {Zurbuchen}, T.~H., \& {Fisk}, L.~A. 2012, \apj, 761, 48, \dodoi{10.1088/0004-637X/761/1/48}

\bibitem[{{Leer} \& {Holzer}(1980)}]{Leer-1980}
{Leer}, E., \& {Holzer}, T.~E. 1980, \jgr, 85, 4681, \dodoi{10.1029/JA085iA09p04681}

\bibitem[{{Lemen} {et~al.}(2012){Lemen}, {Title}, {Akin}, {Boerner}, {Chou}, {Drake}, {Duncan}, {Edwards}, {Friedlaender}, {Heyman}, {Hurlburt}, {Katz}, {Kushner}, {Levay}, {Lindgren}, {Mathur}, {McFeaters}, {Mitchell}, {Rehse}, {Schrijver}, {Springer}, {Stern}, {Tarbell}, {Wuelser}, {Wolfson}, {Yanari}, {Bookbinder}, {Cheimets}, {Caldwell}, {Deluca}, {Gates}, {Golub}, {Park}, {Podgorski}, {Bush}, {Scherrer}, {Gummin}, {Smith}, {Auker}, {Jerram}, {Pool}, {Soufli}, {Windt}, {Beardsley}, {Clapp}, {Lang}, \& {Waltham}}]{Lemen-2012}
{Lemen}, J.~R., {Title}, A.~M., {Akin}, D.~J., {et~al.} 2012, \solphys, 275, 17, \dodoi{10.1007/s11207-011-9776-8}

\bibitem[{{Liewer} {et~al.}(2003){Liewer}, {Neugebauer}, \& {Zurbuchen}}]{Liewer-2003}
{Liewer}, P.~C., {Neugebauer}, M., \& {Zurbuchen}, T. 2003, in American Institute of Physics Conference Series, Vol. 679, Solar Wind Ten, ed. M.~{Velli}, R.~{Bruno}, F.~{Malara}, \& B.~{Bucci}, 51--54, \dodoi{10.1063/1.1618539}

\bibitem[{Liewer {et~al.}(2023)Liewer, Vourlidas, Stenborg, Howard, Qiu, Penteado, Panasenco, \& Braga}]{Liewer-2023}
Liewer, P.~C., Vourlidas, A., Stenborg, G., {et~al.} 2023, The Astrophysical Journal, 948, 24, \dodoi{10.3847/1538-4357/acc8c7}

\bibitem[{Lionello {et~al.}(2001)Lionello, Linker, \& Miki{\'c}}]{Lionello-2001}
Lionello, R., Linker, J.~A., \& Miki{\'c}, Z. 2001, The Astrophysical Journal, 546, 542, \dodoi{10.1086/318254}

\bibitem[{Lionello {et~al.}(2008)Lionello, Linker, \& Miki{\'c}}]{Lionello-2009}
---. 2008, The Astrophysical Journal, 690, 902, \dodoi{10.1088/0004-637X/690/1/902}

\bibitem[{{Lionello} {et~al.}(2005){Lionello}, {Riley}, {Linker}, \& {Miki{\'c}}}]{Lionello-2005}
{Lionello}, R., {Riley}, P., {Linker}, J.~A., \& {Miki{\'c}}, Z. 2005, \apj, 625, 463, \dodoi{10.1086/429268}

\bibitem[{Livi {et~al.}(2022)Livi, Larson, Kasper, Abiad, Case, Klein, Curtis, Dalton, Stevens, Korreck, Ho, Robinson, Tiu, Whittlesey, Verniero, Halekas, McFadden, Marckwordt, Slagle, Abatcha, Rahmati, \& McManus}]{Livi-2022}
Livi, R., Larson, D.~E., Kasper, J.~C., {et~al.} 2022, The Astrophysical Journal, 938, 138, \dodoi{10.3847/1538-4357/ac93f5}

\bibitem[{{Livi} {et~al.}(2023){Livi}, {Lepri}, {Raines}, {Dewey}, {Galvin}, {Louarn}, {Collier}, {Allegrini}, {Alterman}, {Bert}, {Bruno}, {Chornay}, {D'Amicis}, {Eddy}, {Ellis}, {Fauchon-Jones}, {Fedorov}, {Gershkovich}, {Holmes}, {Horbury}, {Kistler}, {Kucharek}, {Lugaz}, {Nieves-Chinchilla}, {O'Brien}, {Ogasawara}, {Owen}, {Phillips}, {Ploof}, {Rivera}, {Spitzer}, {Stubbs}, \& {Wurz}}]{Livi-2023}
{Livi}, S., {Lepri}, S.~T., {Raines}, J.~M., {et~al.} 2023, \aap, 676, A36, \dodoi{10.1051/0004-6361/202346304}

\bibitem[{{Lopez} \& {Freeman}(1986)}]{Lopez-1986}
{Lopez}, R.~E., \& {Freeman}, J.~W. 1986, \jgr, 91, 1701, \dodoi{10.1029/JA091iA02p01701}

\bibitem[{{Macneil} {et~al.}(2022){Macneil}, {Owens}, {Finley}, \& {Matt}}]{Macneil-2022}
{Macneil}, A.~R., {Owens}, M.~J., {Finley}, A.~J., \& {Matt}, S.~P. 2022, \mnras, 509, 2390, \dodoi{10.1093/mnras/stab2965}

\bibitem[{{McComas} {et~al.}(2008){McComas}, {Ebert}, {Elliott}, {Goldstein}, {Gosling}, {Schwadron}, \& {Skoug}}]{McComas-2008}
{McComas}, D.~J., {Ebert}, R.~W., {Elliott}, H.~A., {et~al.} 2008, \grl, 35, L18103, \dodoi{10.1029/2008GL034896}

\bibitem[{{McComas} {et~al.}(2001){McComas}, {Goldstein}, {Gosling}, \& {Skoug}}]{McComas-2001}
{McComas}, D.~J., {Goldstein}, R., {Gosling}, J.~T., \& {Skoug}, R.~M. 2001, \ssr, 97, 99, \dodoi{10.1023/A:1011826111330}

\bibitem[{{McComas} {et~al.}(1998){McComas}, {Riley}, {Gosling}, {Balogh}, \& {Forsyth}}]{McComas-1998}
{McComas}, D.~J., {Riley}, P., {Gosling}, J.~T., {Balogh}, A., \& {Forsyth}, R. 1998, \jgr, 103, 1955, \dodoi{10.1029/97JA01459}

\bibitem[{{Meyer}(1985)}]{Meyer-1985}
{Meyer}, J.~P. 1985, \apjs, 57, 151, \dodoi{10.1086/191000}

\bibitem[{Mostafavi {et~al.}(2022)Mostafavi, Allen, McManus, Ho, Raouafi, Larson, Kasper, \& Bale}]{Mostafavi-2022}
Mostafavi, P., Allen, R.~C., McManus, M.~D., {et~al.} 2022, The Astrophysical Journal Letters, 926, L38, \dodoi{10.3847/2041-8213/ac51e1}

\bibitem[{{M\"uller, D.} {et~al.}(2020){M\"uller, D.}, {St. Cyr, O. C.}, {Zouganelis, I.}, {Gilbert, H. R.}, {Marsden, R.}, {Nieves-Chinchilla, T.}, {Antonucci, E.}, {Auch\`ere, F.}, {Berghmans, D.}, {Horbury, T. S.}, {Howard, R. A.}, {Krucker, S.}, {Maksimovic, M.}, {Owen, C. J.}, {Rochus, P.}, {Rodriguez-Pacheco, J.}, {Romoli, M.}, {Solanki, S. K.}, {Bruno, R.}, {Carlsson, M.}, {Fludra, A.}, {Harra, L.}, {Hassler, D. M.}, {Livi, S.}, {Louarn, P.}, {Peter, H.}, {Sch\"uhle, U.}, {Teriaca, L.}, {del Toro Iniesta, J. C.}, {Wimmer-Schweingruber, R. F.}, {Marsch, E.}, {Velli, M.}, {De Groof, A.}, {Walsh, A.}, \& {Williams, D.}}]{Muller-2020}
{M\"uller, D.}, {St. Cyr, O. C.}, {Zouganelis, I.}, {et~al.} 2020, A\&A, 642, A1, \dodoi{10.1051/0004-6361/202038467}

\bibitem[{{Nolte} \& {Roelof}(1973)}]{Nolte-1973}
{Nolte}, J.~T., \& {Roelof}, E.~C. 1973, \solphys, 33, 241, \dodoi{10.1007/BF00152395}

\bibitem[{{Nolte} {et~al.}(1976){Nolte}, {Krieger}, {Timothy}, {Gold}, {Roelof}, {Vaiana}, {Lazarus}, {Sullivan}, \& {McIntosh}}]{Nolte-1976}
{Nolte}, J.~T., {Krieger}, A.~S., {Timothy}, A.~F., {et~al.} 1976, \solphys, 46, 303, \dodoi{10.1007/BF00149859}

\bibitem[{Ohmi {et~al.}(2004)Ohmi, Kojima, Tokumaru, Fujiki, \& Hakamada}]{Ohmi-2004}
Ohmi, T., Kojima, M., Tokumaru, M., Fujiki, K., \& Hakamada, K. 2004, Advances in Space Research, 33, 689, \dodoi{https://doi.org/10.1016/S0273-1177(03)00238-2}

\bibitem[{{Owen} {et~al.}(2020){Owen}, {Bruno}, {Livi}, {Louarn}, {Al Janabi}, {Allegrini}, {Amoros}, {Baruah}, {Barthe}, {Berthomier}, {Bordon}, {Brockley-Blatt}, {Brysbaert}, {Capuano}, {Collier}, {DeMarco}, {Fedorov}, {Ford}, {Fortunato}, {Fratter}, {Galvin}, {Hancock}, {Heirtzler}, {Kataria}, {Kistler}, {Lepri}, {Lewis}, {Loeffler}, {Marty}, {Mathon}, {Mayall}, {Mele}, {Ogasawara}, {Orlandi}, {Pacros}, {Penou}, {Persyn}, {Petiot}, {Phillips}, {P{\v{r}}ech}, {Raines}, {Reden}, {Rouillard}, {Rousseau}, {Rubiella}, {Seran}, {Spencer}, {Thomas}, {Trevino}, {Verscharen}, {Wurz}, {Alapide}, {Amoruso}, {Andr{\'e}}, {Anekallu}, {Arciuli}, {Arnett}, {Ascolese}, {Bancroft}, {Bland}, {Brysch}, {Calvanese}, {Castronuovo}, {{\v{C}}erm{\'a}k}, {Chornay}, {Clemens}, {Coker}, {Collinson}, {D'Amicis}, {Dandouras}, {Darnley}, {Davies}, {Davison}, {De Los Santos}, {Devoto}, {Dirks}, {Edlund}, {Fazakerley}, {Ferris}, {Frost}, {Fruit}, {Garat}, {G{\'e}not}, {Gibson}, {Gilbert}, {de Giosa}, {Gradone}, {Hailey}, {Horbury},
  {Hunt}, {Jacquey}, {Johnson}, {Lavraud}, {Lawrenson}, {Leblanc}, {Lockhart}, {Maksimovic}, {Malpus}, {Marcucci}, {Mazelle}, {Monti}, {Myers}, {Nguyen}, {Rodriguez-Pacheco}, {Phillips}, {Popecki}, {Rees}, {Rogacki}, {Ruane}, {Rust}, {Salatti}, {Sauvaud}, {Stakhiv}, {Stange}, {Stubbs}, {Taylor}, {Techer}, {Terrier}, {Thibodeaux}, {Urdiales}, {Varsani}, {Walsh}, {Watson}, {Wheeler}, {Willis}, {Wimmer-Schweingruber}, {Winter}, {Yardley}, \& {Zouganelis}}]{Owen-2020}
{Owen}, C.~J., {Bruno}, R., {Livi}, S., {et~al.} 2020, \aap, 642, A16, \dodoi{10.1051/0004-6361/201937259}

\bibitem[{Owens(2018)}]{Owens-2018}
Owens, M.~J. 2018, Solar Physics, 293, 122, \dodoi{10.1007/s11207-018-1343-0}

\bibitem[{{Owocki} {et~al.}(1983){Owocki}, {Holzer}, \& {Hundhausen}}]{Owocki-1983}
{Owocki}, S.~P., {Holzer}, T.~E., \& {Hundhausen}, A.~J. 1983, \apj, 275, 354, \dodoi{10.1086/161538}

\bibitem[{{Panasenco} {et~al.}(2013){Panasenco}, {Martin}, {Velli}, \& {Vourlidas}}]{Panasenco-2013}
{Panasenco}, O., {Martin}, S.~F., {Velli}, M., \& {Vourlidas}, A. 2013, \solphys, 287, 391, \dodoi{10.1007/s11207-012-0194-3}

\bibitem[{{Panasenco} \& {Velli}(2013)}]{PV-2013}
{Panasenco}, O., \& {Velli}, M. 2013, in American Institute of Physics Conference Series, Vol. 1539, Solar Wind 13, ed. G.~P. {Zank}, J.~{Borovsky}, R.~{Bruno}, J.~{Cirtain}, S.~{Cranmer}, H.~{Elliott}, J.~{Giacalone}, W.~{Gonzalez}, G.~{Li}, E.~{Marsch}, E.~{Moebius}, N.~{Pogorelov}, J.~{Spann}, \& O.~{Verkhoglyadova}, 50--53, \dodoi{10.1063/1.4810987}

\bibitem[{{Panasenco} {et~al.}(2019){Panasenco}, {Velli}, \& {Panasenco}}]{Panasenco-2019}
{Panasenco}, O., {Velli}, M., \& {Panasenco}, A. 2019, \apj, 873, 25, \dodoi{10.3847/1538-4357/ab017c}

\bibitem[{{Parker}(1958)}]{Parker-1958}
{Parker}, E.~N. 1958, \apj, 128, 664, \dodoi{10.1086/146579}

\bibitem[{{Perrone} {et~al.}(2020){Perrone}, {D'Amicis}, {De Marco}, {Matteini}, {Stansby}, {Bruno}, \& {Horbury}}]{Perrone-2020}
{Perrone}, D., {D'Amicis}, R., {De Marco}, R., {et~al.} 2020, \aap, 633, A166, \dodoi{10.1051/0004-6361/201937064}

\bibitem[{{Pesnell} {et~al.}(2012){Pesnell}, {Thompson}, \& {Chamberlin}}]{Pesnell-2012}
{Pesnell}, W.~D., {Thompson}, B.~J., \& {Chamberlin}, P.~C. 2012, \solphys, 275, 3, \dodoi{10.1007/s11207-011-9841-3}

\bibitem[{{Pulupa} {et~al.}(2017){Pulupa}, {Bale}, {Bonnell}, {Bowen}, {Carruth}, {Goetz}, {Gordon}, {Harvey}, {Maksimovic}, {Mart{\'\i}nez-Oliveros}, {Moncuquet}, {Saint-Hilaire}, {Seitz}, \& {Sundkvist}}]{Pulupa-2016}
{Pulupa}, M., {Bale}, S.~D., {Bonnell}, J.~W., {et~al.} 2017, Journal of Geophysical Research (Space Physics), 122, 2836, \dodoi{10.1002/2016JA023345}

\bibitem[{{Riley}(2021)}]{psipy}
{Riley}, P. 2021, EDP Sciences for European Southern Observatory.

\bibitem[{{Riley} {et~al.}(2006){Riley}, {Linker}, {Miki{\'c}}, {Lionello}, {Ledvina}, \& {Luhmann}}]{Riley-2006}
{Riley}, P., {Linker}, J.~A., {Miki{\'c}}, Z., {et~al.} 2006, \apj, 653, 1510, \dodoi{10.1086/508565}

\bibitem[{{Riley} {et~al.}(2021){Riley}, {Lionello}, {Caplan}, {Downs}, {Linker}, {Badman}, \& {Stevens}}]{Riley-2021}
{Riley}, P., {Lionello}, R., {Caplan}, R.~M., {et~al.} 2021, \aap, 650, A19, \dodoi{10.1051/0004-6361/202039815}

\bibitem[{{Rivera} {et~al.}(Submitted){Rivera}, {Badman}, \& {Stevens}}]{Rivera-2023}
{Rivera}, Y.~J., {Badman}, S.~T., \& {Stevens}, M.~L. Submitted, Science

\bibitem[{Romeo {et~al.}(2023)Romeo, Braga, Badman, Larson, Stevens, Huang, Phan, Rahmati, Livi, Alnussirat, Whittlesey, Szabo, Klein, Niembro-Hernandez, Paulson, Verniero, Lario, Raouafi, Ervin, Kasper, Pulupa, Bale, \& Linton}]{Romeo-2023}
Romeo, O.~M., Braga, C.~R., Badman, S.~T., {et~al.} 2023, The Astrophysical Journal, 954, 168, \dodoi{10.3847/1538-4357/ace62e}

\bibitem[{{Rouillard} {et~al.}(2010){Rouillard}, {Davies}, {Lavraud}, {Forsyth}, {Savani}, {Bewsher}, {Brown}, {Sheeley}, {Davis}, {Harrison}, {Howard}, {Vourlidas}, {Lockwood}, {Crothers}, \& {Eyles}}]{Rouillard-2010}
{Rouillard}, A.~P., {Davies}, J.~A., {Lavraud}, B., {et~al.} 2010, Journal of Geophysical Research (Space Physics), 115, A04103, \dodoi{10.1029/2009JA014471}

\bibitem[{{Schatten} {et~al.}(1969){Schatten}, {Wilcox}, \& {Ness}}]{Schatten-1969}
{Schatten}, K.~H., {Wilcox}, J.~M., \& {Ness}, N.~F. 1969, \solphys, 6, 442, \dodoi{10.1007/BF00146478}

\bibitem[{{Scherrer} {et~al.}(2012){Scherrer}, {Schou}, {Bush}, {Kosovichev}, {Bogart}, {Hoeksema}, {Liu}, {Duvall}, {Zhao}, {Title}, {Schrijver}, {Tarbell}, \& {Tomczyk}}]{Scherrer-2012}
{Scherrer}, P.~H., {Schou}, J., {Bush}, R.~I., {et~al.} 2012, \solphys, 275, 207, \dodoi{10.1007/s11207-011-9834-2}

\bibitem[{{Schwenn}(2006)}]{Schwenn-2006}
{Schwenn}, R. 2006, \ssr, 124, 51, \dodoi{10.1007/s11214-006-9099-5}

\bibitem[{{Sheeley} {et~al.}(2009){Sheeley}, {Lee}, {Casto}, {Wang}, \& {Rich}}]{Sheeley-2009}
{Sheeley}, N.~R., J., {Lee}, D.~D.~H., {Casto}, K.~P., {Wang}, Y.~M., \& {Rich}, N.~B. 2009, \apj, 694, 1471, \dodoi{10.1088/0004-637X/694/2/1471}

\bibitem[{{Sheeley} {et~al.}(2013){Sheeley}, {Martin}, {Panasenco}, \& {Warren}}]{Sheeley-2013}
{Sheeley}, N.~R., J., {Martin}, S.~F., {Panasenco}, O., \& {Warren}, H.~P. 2013, \apj, 772, 88, \dodoi{10.1088/0004-637X/772/2/88}

\bibitem[{{Smith} {et~al.}(1978){Smith}, {Tsurutani}, \& {Rosenberg}}]{Smith-1978}
{Smith}, E.~J., {Tsurutani}, B.~T., \& {Rosenberg}, R.~L. 1978, \jgr, 83, 717, \dodoi{10.1029/JA083iA02p00717}

\bibitem[{Snyder \& Neugebauer(1966)}]{Snyder-1966}
Snyder, C.~W., \& Neugebauer, M. 1966, The Solar Wind (Pergamon Press)

\bibitem[{{Stakhiv} {et~al.}(2015){Stakhiv}, {Landi}, {Lepri}, {Oran}, \& {Zurbuchen}}]{Stakhiv-2015}
{Stakhiv}, M., {Landi}, E., {Lepri}, S.~T., {Oran}, R., \& {Zurbuchen}, T.~H. 2015, \apj, 801, 100, \dodoi{10.1088/0004-637X/801/2/100}

\bibitem[{{Stansby} {et~al.}(2020{\natexlab{a}}){Stansby}, {Baker}, {Brooks}, \& {Owen}}]{Stansby-2020comp}
{Stansby}, D., {Baker}, D., {Brooks}, D.~H., \& {Owen}, C.~J. 2020{\natexlab{a}}, \aap, 640, A28, \dodoi{10.1051/0004-6361/202038319}

\bibitem[{Stansby {et~al.}(2018)Stansby, Horbury, \& Matteini}]{Stansby-2018}
Stansby, D., Horbury, T.~S., \& Matteini, L. 2018, Monthly Notices of the Royal Astronomical Society, 482, 1706, \dodoi{10.1093/mnras/sty2814}

\bibitem[{{Stansby} {et~al.}(2020{\natexlab{b}}){Stansby}, {Matteini}, {Horbury}, {Perrone}, {D'Amicis}, \& {Ber{\v{c}}i{\v{c}}}}]{Stansby-2020alf}
{Stansby}, D., {Matteini}, L., {Horbury}, T.~S., {et~al.} 2020{\natexlab{b}}, \mnras, 492, 39, \dodoi{10.1093/mnras/stz3422}

\bibitem[{Stansby {et~al.}(2020)Stansby, Yeates, \& Badman}]{pfss}
Stansby, D., Yeates, A., \& Badman, S. 2020, Journal of Open Source Software, 5, 2732, \dodoi{10.21105/joss.02732}

\bibitem[{Stansby {et~al.}(2022)Stansby, Rai, Argall, JeffreyBroll, Haythornthwaite, Teunissen, Shaw, xypnox, Saha, Ireland, Lim, Badman, Mishra, Badger, dupuisIRT, \& tlml}]{heliopy}
Stansby, D., Rai, Y., Argall, M., {et~al.} 2022

\bibitem[{{Suess} {et~al.}(2009){Suess}, {Ko}, {von Steiger}, \& {Moore}}]{Suess-2009}
{Suess}, S.~T., {Ko}, Y.~K., {von Steiger}, R., \& {Moore}, R.~L. 2009, Journal of Geophysical Research (Space Physics), 114, A04103, \dodoi{10.1029/2008JA013704}

\bibitem[{{Tu} \& {Marsch}(1995)}]{Tu-1995}
{Tu}, C.~Y., \& {Marsch}, E. 1995, \ssr, 73, 1, \dodoi{10.1007/BF00748891}

\bibitem[{{Viall} \& {Borovsky}(2020)}]{Viall-2020}
{Viall}, N.~M., \& {Borovsky}, J.~E. 2020, Journal of Geophysical Research (Space Physics), 125, e26005, \dodoi{10.1029/2018JA026005}

\bibitem[{Virtanen {et~al.}(2020)Virtanen, Gommers, Oliphant, Haberland, Reddy, Cournapeau, Burovski, Peterson, Weckesser, Bright, van~der Walt, Brett, Wilson, Millman, Mayorov, Nelson, Jones, Kern, Larson, Carey, Polat, Feng, Moore, VanderPlas, Laxalde, Perktold, Cimrman, Henriksen, Quintero, Harris, Archibald, Ribeiro, Pedregosa, van Mulbregt, Vijaykumar, Bardelli, Rothberg, Hilboll, Kloeckner, Scopatz, Lee, Rokem, Woods, Fulton, Masson, H{\"a}ggstr{\"o}m, Fitzgerald, Nicholson, Hagen, Pasechnik, Olivetti, Martin, Wieser, Silva, Lenders, Wilhelm, Young, Price, Ingold, Allen, Lee, Audren, Probst, Dietrich, Silterra, Webber, Slavi{\v c}, Nothman, Buchner, Kulick, Sch{\"o}nberger, de~Miranda~Cardoso, Reimer, Harrington, Rodr{\'\i}guez, Nunez-Iglesias, Kuczynski, Tritz, Thoma, Newville, K{\"u}mmerer, Bolingbroke, Tartre, Pak, Smith, Nowaczyk, Shebanov, Pavlyk, Brodtkorb, Lee, McGibbon, Feldbauer, Lewis, Tygier, Sievert, Vigna, Peterson, More, Pudlik, Oshima, Pingel, Robitaille, Spura, Jones, Cera, Leslie, Zito,
  Krauss, Upadhyay, Halchenko, V{\'a}zquez-Baeza, \& Contributors}]{scipy}
Virtanen, P., Gommers, R., Oliphant, T.~E., {et~al.} 2020, Nature Methods, 17, 261, \dodoi{10.1038/s41592-019-0686-2}

\bibitem[{{von Steiger} \& {Zurbuchen}(2011)}]{vonSteiger-2011}
{von Steiger}, R., \& {Zurbuchen}, T.~H. 2011, Journal of Geophysical Research (Space Physics), 116, A01105, \dodoi{10.1029/2010JA015835}

\bibitem[{von Steiger {et~al.}(2000)von Steiger, Schwadron, Fisk, Geiss, Gloeckler, Hefti, Wilken, Wimmer-Schweingruber, \& Zurbuchen}]{vonSteiger-2000}
von Steiger, R., Schwadron, N.~A., Fisk, L.~A., {et~al.} 2000, Journal of Geophysical Research: Space Physics, 105, 27217, \dodoi{https://doi.org/10.1029/1999JA000358}

\bibitem[{{Wang}(2009)}]{Wang-2009}
{Wang}, Y.~M. 2009, \ssr, 144, 383, \dodoi{10.1007/s11214-008-9434-0}

\bibitem[{{Wang}(2013)}]{Wang-2013}
---. 2013, \apjl, 775, L46, \dodoi{10.1088/2041-8205/775/2/L46}

\bibitem[{{Wang}(2016)}]{Wang-2016}
---. 2016, \apj, 833, 121, \dodoi{10.3847/1538-4357/833/1/121}

\bibitem[{{Wang} {et~al.}(2012){Wang}, {Grappin}, {Robbrecht}, \& {Sheeley}}]{Wang-2012}
{Wang}, Y.~M., {Grappin}, R., {Robbrecht}, E., \& {Sheeley}, N.~R., J. 2012, \apj, 749, 182, \dodoi{10.1088/0004-637X/749/2/182}

\bibitem[{Wang {et~al.}(2009)Wang, Ko, \& Grappin}]{Wang-2009a}
Wang, Y.-M., Ko, Y.-K., \& Grappin, R. 2009, The Astrophysical Journal, 691, 760, \dodoi{10.1088/0004-637X/691/1/760}

\bibitem[{{Wang} \& {Sheeley}(1990)}]{Wang-1990}
{Wang}, Y.~M., \& {Sheeley}, N.~R., J. 1990, \apj, 355, 726, \dodoi{10.1086/168805}

\bibitem[{{Wang} \& {Sheeley}(1997)}]{Wang-1997}
---. 1997, \grl, 24, 3141, \dodoi{10.1029/97GL53305}

\bibitem[{{Wang} {et~al.}(2000){Wang}, {Sheeley}, {Socker}, {Howard}, \& {Rich}}]{Wang-2000b}
{Wang}, Y.~M., {Sheeley}, N.~R., {Socker}, D.~G., {Howard}, R.~A., \& {Rich}, N.~B. 2000, \jgr, 105, 25133, \dodoi{10.1029/2000JA000149}

\bibitem[{{W}es {M}c{K}inney(2010)}]{pandas}
{W}es {M}c{K}inney. 2010, in {P}roceedings of the 9th {P}ython in {S}cience {C}onference, ed. {S}t\'efan van~der {W}alt \& {J}arrod {M}illman, 56 -- 61, \dodoi{10.25080/Majora-92bf1922-00a}

\bibitem[{Whittlesey {et~al.}(2020)Whittlesey, Larson, Kasper, Halekas, Abatcha, Abiad, Berthomier, Case, Chen, Curtis, Dalton, Klein, Korreck, Livi, Ludlam, Marckwordt, Rahmati, Robinson, Slagle, Stevens, Tiu, \& Verniero}]{Whittlesey-2020}
Whittlesey, P.~L., Larson, D.~E., Kasper, J.~C., {et~al.} 2020, The Astrophysical Journal Supplement Series, 246, 74, \dodoi{10.3847/1538-4365/ab7370}

\bibitem[{Wicks {et~al.}(2013)Wicks, Roberts, Mallet, Schekochihin, Horbury, \& Chen}]{Wicks-2013}
Wicks, R.~T., Roberts, D.~A., Mallet, A., {et~al.} 2013, The Astrophysical Journal, 778, 177, \dodoi{10.1088/0004-637X/778/2/177}

\bibitem[{{Widing} \& {Feldman}(2001)}]{Widing-2001}
{Widing}, K.~G., \& {Feldman}, U. 2001, \apj, 555, 426, \dodoi{10.1086/321482}

\bibitem[{Worden \& Harvey(2000)}]{Worden-2000}
Worden, J., \& Harvey, J. 2000, Solar Physics, 195, 247, \dodoi{10.1023/A:1005272502885}

\bibitem[{{Zhao} \& {Fisk}(2011)}]{Zhao-2011}
{Zhao}, L., \& {Fisk}, L. 2011, \solphys, 274, 379, \dodoi{10.1007/s11207-011-9840-4}

\bibitem[{Zhao {et~al.}(2017)Zhao, Landi, Lepri, Kocher, Zurbuchen, Fisk, \& Raines}]{Zhao-2017}
Zhao, L., Landi, E., Lepri, S.~T., {et~al.} 2017, The Astrophysical Journal Supplement Series, 228, 4, \dodoi{10.3847/1538-4365/228/1/4}

\bibitem[{{Zhao} {et~al.}(2014){Zhao}, {Landi}, {Zurbuchen}, {Fisk}, \& {Lepri}}]{Zhao-2014}
{Zhao}, L., {Landi}, E., {Zurbuchen}, T.~H., {Fisk}, L.~A., \& {Lepri}, S.~T. 2014, \apj, 793, 44, \dodoi{10.1088/0004-637X/793/1/44}

\bibitem[{{Zhao} {et~al.}(2009){Zhao}, {Zurbuchen}, \& {Fisk}}]{Zhao-2009}
{Zhao}, L., {Zurbuchen}, T.~H., \& {Fisk}, L.~A. 2009, \grl, 36, L14104, \dodoi{10.1029/2009GL039181}

\bibitem[{{Zurbuchen}(2007)}]{Zurbuchen-2007}
{Zurbuchen}, T.~H. 2007, \araa, 45, 297, \dodoi{10.1146/annurev.astro.45.010807.154030}

\end{thebibliography}
\bibliographystyle{aasjournal}

\end{document}